\documentclass[twocolumn]{aastex62}

\newcommand{\arcs}{$^{\prime\prime}$} 
\newcommand{\beq}{\begin{equation}\begin{aligned}}
\newcommand{\eeq}{\end{aligned}\end{equation}}

\newcommand{\msun}{M$_\odot$}

\usepackage{scalerel,amssymb}
\usepackage{amsmath}
\usepackage{enumitem}
\usepackage[]{url}
\usepackage{lineno}

\accepted{\today}
\submitjournal{ApJ}

\shorttitle{Structures of Dwarf Satellites of MW-like Galaxies}
\shortauthors{Carlsten et al.}

\begin{document}

\title{ELVES I: Structures of Dwarf Satellites of MW-like Galaxies; Morphology, Scaling Relations, and Intrinsic Shapes}

\correspondingauthor{Scott G. Carlsten}
\email{scottgc@princeton.edu}

\author[0000-0002-5382-2898]{Scott G. Carlsten}
\affil{Department of Astrophysical Sciences, 4 Ivy Lane, Princeton University, Princeton, NJ 08544}

\author[0000-0002-5612-3427]{Jenny E. Greene}
\affil{Department of Astrophysical Sciences, 4 Ivy Lane, Princeton University, Princeton, NJ 08544}

\author[0000-0003-4970-2874]{Johnny P. Greco}
\altaffiliation{NSF Astronomy \& Astrophysics Postdoctoral Fellow}
\affiliation{Center for Cosmology and AstroParticle Physics (CCAPP), The Ohio State University, Columbus, OH 43210, USA}

\author[0000-0002-1691-8217]{Rachael L. Beaton}
\altaffiliation{Hubble Fellow}
\affiliation{Department of Astrophysical Sciences, 4 Ivy Lane, Princeton University, Princeton, NJ 08544}
\affiliation{The Observatories of the Carnegie Institution for Science, 813 Santa Barbara St., Pasadena, CA~91101\\}

\author[0000-0002-0332-177X]{Erin Kado-Fong}
\affil{Department of Astrophysical Sciences, 4 Ivy Lane, Princeton University, Princeton, NJ 08544}

\begin{abstract}
The structure of a dwarf galaxy is an important probe into the effects of stellar feedback and environment. Using an unprecedented sample of 223 low-mass satellites from the ongoing Exploration of Local VolumE Satellites (ELVES) Survey, we explore the structures of dwarf satellites in the mass range $10^{5.5}<M_\star<10^{8.5}$ \msun. We survey satellites around 80\% of the massive, $M_K<-22.4$ mag, hosts in the Local Volume. Our sample of dwarf satellites is complete to luminosities of $M_V<-9$ mag and surface brightness $\mu_{0,V}<26.5$ mag arcsec$^{-2}$ within at least $\sim200$ projected kpc. We separate the satellites into late- and early-type, finding the  mass-size  relations  are  very  similar  between them, to within $\sim5$\%.  This similarity indicates that the quenching and transformation of a late-type dwarf into an early-type involves only very mild size evolution. Considering the distribution of apparent ellipticities, we infer the intrinsic shapes of the early- and late-type samples. Combining with literature samples, we find that both types of dwarfs get thicker at  fainter luminosities but early-types are always rounder at fixed luminosity. Finally, we compare the LV satellites with dwarf samples from the cores of the Virgo and Fornax clusters. We find that the cluster satellites show similar scaling relations to the LV early-type dwarfs, but are roughly $10\%$ larger at fixed mass, which we interpret as being due to tidal heating in the cluster environments. The dwarf structure results presented here are a useful reference for simulations of dwarf galaxy formation and the transformation of dwarf irregulars into spheroidals.

\end{abstract}


\section{Introduction}
Low-mass ($M_\star < 10^9$ \msun) galaxies are important windows into a variety of astrophysical processes. Their occupation of low-mass dark matter halos, and the fact that they are generally very dark matter dominated, makes them important probes into the physics of dark matter. Their shallow potential wells makes them sensitive to the physics of star formation feedback, and their relatively simple formation histories make it easier to delineate the effects of different aspects of galaxy formation than for massive galaxies. Finally, dwarf galaxies that are satellites of more massive galaxies demonstrate the various ways that galaxies can be affected by their environments. 

The photometric appearance of a galaxy, including its morphology and structural parameters, is one of the simplest observables to measure yet it encodes significant information on the formation and evolution of the galaxy \citep[e.g.][]{binggeli1987, lisker2007, rsj2010, rsj2016, janz2016}. Additionally, the scaling relations between various photometric and kinematic properties of galaxies have long been used to elucidate the formation pathways and relevant physics for different classes of galaxies \citep[e.g.][]{faber1976, kormendy1977, djorgovski1987, graham2003, shen2003, boselli2008_scaling, kormendy2009, misgeld2011, eigenthaler2018}. In particular, dwarf galaxies are observed to have scaling relations between their luminosity (or stellar mass), effective radii, and surface brightness (central or effective). The physical origins of these correlations, and hence what physics determines the observed structure of dwarf galaxies, is a continual area of interest for galaxy formation. In particular, the structure of a dwarf galaxy should reflect the effects of at least both stellar feedback and the environment the dwarf lives in.

The idea that stellar feedback (particularly via supernovae) can have a large impact on the ISM and CGM of dwarf galaxies has been around for decades \citep[e.g.][]{larson1974, dekel1986}. The modern generation of hydrodynamic galaxy formation simulations have expanded on these early ideas, and many simulation projects predict that feedback will lead to repetitive cycles of star-formation and gas blowout (``bursty star-formation") \citep[e.g.][]{stinson2009, muratov2015, christensen2016}. While much of the research in this area has focused on how these bursts affect the underlying collisionless dark matter (DM) density profile (``cusp-to-core" transformation) \citep{mashchenko2008, governato2010, pontzen2012}, it was realized early on that these repetitive outflows could also affect the distribution of stars \citep{read2005, read2006, stinson2009, maxwell2012, teyssier2013, elbadry2016, dicintio2017, chan2018}, increasing the size of stellar orbits and even creating stellar halos for dwarfs. Recent hydrodynamic simulation projects have simulated large samples of dwarfs (both in the field and as satellites) and made testable predictions for the structural scaling relations \cite[e.g.][]{martin2019, liao2019, jackson2020, wright2020, applebaum2020}.

The understanding that the morphology and structure of a dwarf galaxy is related to its environment also is not new \citep{dressler1980, binggeli1987}. Specifically in the context of satellite dwarfs of MW-mass hosts, the environment will have two main effects on the structure of low-mass dwarfs. First, the hot gas halo of the host can remove the gas reservoirs of the dwarfs quickly via ram pressure stripping \citep{gunn1972, grebel2003, boselli2008, fillingham2015} or at least cutoff the supply of new gas to the dwarfs. This explains the observed population of predominantly quenched and gas-free satellite dSph's in the Local Group (LG) \citep{grcevich2009, spekkens2014}. Once quenched, the surface brightness of the dwarfs will slowly fade as the stellar populations in the dwarf passively evolve. 

The second main environmental effect comes from the tidal field of the host. This includes both any initial morphological transformation from a gas-rich dwarf irregular to a dwarf spheroidal (dSph) and the subsequent evolution of the dSph as it orbits in the halo of its host. The most widely written about model for the initial transformation is that of `tidal stirring' of initially gas-rich rotationally supported field dIrrs that fall into the halo of the host \citep{mayer2001a, mayer2001b, mayer2006, kazantzidis2011, kazantzidis2013, kazantzidis2017}. This process generally relies on the formation of disk instabilities, particularly the formation of a bar, over repeated pericentric passages to facilitate the loss of angular momentum from the dwarf and the formation of a dispersion supported system. Structurally, this process will change the dwarf from a fairly thin disk to a mildly oblate spheroid. However, recent work has shown that low-mass isolated dIrrs are actually generally quite thick, not flattened, \citep{kaufman2007, rsj2010, roychowdhury2010, roychowdhury2013}, in both stars and gas, and can even be dispersion supported \citep{wheeler2017}, indicating that a drastic morphological change might not even be needed to explain the properties of dSphs. Either way, once the dSph exists, it will continue to evolve in the tidal field of the host via both tidal stripping and tidal heating \citep{jiang2020}. Controlled simulations indicate that the net effect of tidal processing is generally that the satellite's half-light radius will increase, and surface brightness decrease, as it loses mass \citep{penarrubia2008, errani2015, errani2018}, although this depends on the density profile of the underlying dark matter (DM) halo. 

Dwarfs in denser, cluster-like environments will experience the above effects (likely even more intensely) in addition to environmental effects specific to clusters, including exposure to the overall tidal field of the cluster and repeated, high-speed encounters with cluster members \citep[i.e. ``galaxy harrassment'',][]{moore1998, smith2015}. Isolated dwarfs in the field, on the other hand, will not experience any of these affects but will more intensely feel the effect of stellar feedback due to their ubiquitous active star formation \citep[e.g.][]{geha2012}. Unravelling the interrelated effects of stellar feedback and environment on the structural properties of dwarfs clearly requires complete samples of dwarfs across a range of environments.

In this paper, we use a new sample of confirmed dwarf satellites with high-quality photometry around massive, Milky Way-like (MW) and loose group hosts in the Local ($D<12$ Mpc) Volume (LV) with well-understood completeness to explore the structural properties and scaling relations of faint ($M_V<-9$ mag) dwarfs in these environments. Our work extends previous work in this area in two important ways. First, while numerous previous works have considered the structure of dwarfs in the LG, the current sample of dwarfs is an order of magnitude larger, allowing average trends to be determined much more precisely. This is timely as comparable samples have recently been produced for the nearby Virgo and Fornax clusters through the Next Generation Virgo Survey \citep[NGVS][]{ferrarese2012, roediger2017, ferrarese2020, lim2020} and Next Generation Fornax Survey \citep[NGFS;][]{munoz2015, eigenthaler2018, ordenes-briceno2018} \footnote{See also the Fornax Deep Survey \citep{venhola2018}}, allowing for detailed comparison between environments. For the purpose of this work, we use ‘environment’ of dwarf galaxies to largely mean the mass of the parent halo that the dwarf is a satellite of.

Second, the sample of satellites of nearby MW-like galaxies includes a large population of late-type dwarfs presumably consisting primarily of dwarfs that only recently fell into the halos of their hosts and are thus either minimally processed by their environment or experiencing only the first stages of this process. A volume-limited sample of truly isolated dwarfs at these luminosities with well-understood surface brightness completeness currently does not exist\footnote{Such a sample likely will not exist until the surveys of the Vera Rubin Observatory and the Nancy Grace Roman Space Telescope.}. The Catalog of Neighboring Galaxies of \citet{karachentsev2004, karachentsev2013} includes many isolated dwarfs in the Local Volume and includes many dwarfs with TRGB distances based on numerous programs with \emph{HST} \citep[e.g.][]{karachentsev2006, karachentsev2007}. However, the completeness of the catalog (particularly with respect to surface brightness) is not well established. Additionally the reported photometry commonly relies on uncertain photographic plate measurements. The late-type sub-sample of satellites then represents one of the best current ways to understand what a field sample of low-mass dwarfs will look like and will help delineate the effects of stellar feedback and environment.

In Section \ref{sec:data}, we overview the observational sample used in this work. In Section \ref{sec:results} we present the observational results of the dwarf structural scaling relations. In Section \ref{sec:shapes}, we analyze the intrinsic shapes of the dwarf samples. In Section \ref{sec:disc}, we discuss the results as they relate to the current understanding of galaxy formation. Finally, we provide a summary of the main results in Section \ref{sec:concl}.

\section{Galaxy Sample and Photometry}
\label{sec:data}

\subsection{ELVES Sample Overview}
\label{sec:survey}

Surveying and cataloging the satellite systems of nearby MW-like galaxies has seen an explosion of progress in recent years \citep[e.g.][]{karachentsev2015, crnojevic2016, madcash, bennet2017, muller2017, muller2018, geha2017, greco2018, tanaka2018, park2019, byun2020, mao2020, habas2020, davis2021} with many groups using wide-field imaging to catalog a plethora of low luminosity, low surface brightness candidate satellites. The difficulty in this endeavor is in determining the distance to candidate satellites to confirm their association with a host. Since these groups are quite sparse (compared to Virgo or Fornax, for instance), background contaminates can often consist of a majority \citep[$>80\%$, e.g.][]{sbf_m101, carlsten2020a} of candidate satellites selected on their low-surface brightness, diffuse morphology. The inferred physical properties of dwarfs clearly depends on their distance, thus it is crucial we only consider satellite systems that have full (or nearly full) distance information. Additionally, it is important that we consider systems that have well-defined completeness limits, with respect to both luminosity and surface brightness. 

Thus, for our primary observational sample we use results from the ongoing Exploration of Local VolumE Satellites (ELVES) survey to obtain a nearly \textit{volume limited} sample of many well-surveyed satellites systems in the Local ($D<12$ Mpc) Volume. The explicit goal of the survey is to survey the classical-mass satellites down to $M_V\sim-9$ and within 300 kpc projected of all massive, $M_K < -22.4$ mag, hosts in this volume with full or nearly full distance information for the satellites. The details of this host selection and the list of hosts will be presented in a future paper describing the survey in more detail. Satellite candidates are detected using deep, wide-field imaging combined with the detection algorithm specialized for finding low-surface brightness, diffuse dwarf galaxies of \citet{carlsten2020a} and \citet{greco2018}.

The candidate satellites are confirmed with a variety of distance measurements including archival redshifts and TRGB distances, but primarily surface brightness fluctuations (SBF) is used. SBF \citep[e.g.][]{tonry1988, jerjen_field, jerjen_field2, sbf_calib, sbf_m101, greco2020} can produce relatively precise distance errors ($\lesssim15\%$) for LSB dwarfs using modest ground based data. This is enough to confirm the association of candidate satellites or not in most cases\footnote{With this distance precision, we fully expect some of the `confirmed' satellites to be near-field objects that are $\lesssim1$ Mpc from their hosts but still outside the host virial radius. The fraction of these near-field dwarfs is expected to be $\sim10-15\%$ \citep{carlsten2020b, carlsten2020c}.}.

For most of the survey, we are making use of  the DESI Legacy Imaging Surveys \citep{decals}\footnote{\url{https://www.legacysurvey.org/}} which includes both the Beijing-Arizona Sky Survey \citep[BASS;][]{bass1, bass2} and the DECam Legacy Survey (henceforth we refer to all these surveys together as DECaLS). We find we can readily detect dwarfs down to $M_V\sim-9$ (see \S\ref{sec:completeness}) even using these relatively shallow surveys, however the depth and PSF size are not adequate for SBF for most of the low-mass dwarfs considered in this survey. We are then using deeper CFHT/MegaCam, Subaru/HSC, or Gemini/GMOS imaging to measure the dwarf distances via SBF. We include results using Gemini programs FT-2020A-060 and US-2020B-037 (PI: S. Carlsten). The rest of the data used is all archival. For hosts where deeper data is available (i.e. Subaru/HSC), we still do the object detection on the DECaLS data for as much consistency with the other hosts as possible. 

The list of hosts is given in Table \ref{tab:hosts} where we list the different data sets used, number of satellites for each host, and the radial extent of the surveys. For the object detection, we use Data Release 8 of DECaLS but the photometry makes use of the recently released Data Release 9 which features an improved sky subtraction process. More details of the data usage will be given in a future survey overview paper. However, we emphasize that all of the major steps involved have been published. \citet{carlsten2020a} details the detection algorithm and completeness checks, \citep{carlsten2020b} describes the use of SBF on the CFHT data, and \citet{carlsten2020c} details the use of SBF with Gemini data.  Only minor changes have been adapted for use here.

Six more hosts have been surveyed out to at least roughly 150 kpc projected and down to this luminosity limit by other groups in the literature: MW, M31, M81, NGC 5128 (CenA), M94, and M101. The references for the satellite surveys and the sources for the satellite photometry are given in Table \ref{tab:hosts}.


Thus, altogether we include satellites of 24 massive hosts in the LV. 13 of these are complete to 300 kpc projected from their host while the rest are surveyed only out to $\sim150-200$ kpc. Out of the entire LV sample, roughly 3/4 of the hosts are included here.

\begin{deluxetable*}{ccccccc}
\tablecaption{ELVES Host Information\label{tab:hosts}}
\tablewidth{\textwidth}
\tablehead{
\colhead{Name} & \colhead{Dist} & \colhead{N, confirmed}  & \colhead{$r_\mathrm{cover}$} & \colhead{Detection} & \colhead{Confirmation} & \colhead{Photometry} \\ 
\colhead{} & \colhead{Mpc} & \colhead{(unconfirmed)}  & \colhead{kpc} & \colhead{} & \colhead{} & \colhead{}  }
\startdata
MW & 0 & 7(0) & 300 & -- & -- & M12 \\
M31 & 0.78 & 13(0) & 300 & -- & -- & M12 \\
NGC1023 & 10.4 & 12(3) & 300 & CFHT/MegaCam & CFHT/MegaCam & CFHT/MegaCam \\
NGC2903 & 9.0 & 6(1) & 300 & CFHT/MegaCam,DECaLS & CFHT/MegaCam,Gemini & CFHT/MegaCam,DECaLS \\
NGC4258 & 7.2 & 5(0) & 150 & CFHT/MegaCam & CFHT/MegaCam & CFHT/MegaCam \\
NGC4565 & 11.9 & 3(5) & 150 & CFHT/MegaCam & CFHT/MegaCam & CFHT/MegaCam \\
NGC4631 & 7.4 & 10(0) & 200 & DECaLS & CFHT/MegaCam,Gemini & CFHT/MegaCam,DECaLS \\
M51 & 8.58 & 0(2) & 150 & CFHT/MegaCam & CFHT/MegaCam & CFHT/MegaCam \\
M104 & 9.55 & 11(2) & 150 & CFHT/MegaCam & CFHT/MegaCam & CFHT/MegaCam \\
NGC891 & 9.12 & 3(0) & 200 & CFHT/MegaCam & CFHT/MegaCam & CFHT/MegaCam \\
NGC6744 & 8.95 & 5(6) & 200 & DECam & DECam & DECam \\
M101 & 6.5 & 7(0) & 300 & DECaLS & B19,Subaru/HSC & CFHT/MegaCam,DECaLS \\
M94 & 4.2 & 3(3) & 200 & DECaLS & S18 & DECaLS \\
NGC253 & 3.56 & 3(0) & 300 & DECaLS & DECaLS & DECaLS \\
M64 & 5.3 & 8(2) & 300 & DECaLS & DECaLS/HSC & DECaLS \\
NGC5055 & 8.87 & 8(0) & 200 & DECaLS & Subaru/HSC & DECaLS \\
NGC4517 & 8.34 & 7(0) & 300 & DECaLS & Subaru/HSC & DECaLS \\
NGC3627 & 10.5 & 11(5) & 200 & DECaLS & CFHT/MegaCam,Subaru/HSC & DECaLS \\
NGC3379 & 10.7 & 29(16) & 300 & DECaLS & CFHT/MegaCam,Subaru/HSC & DECaLS \\
M81 & 3.61 & 19(0) & 300 & C13 & C13 & DECaLS \\
CENA & 3.66 & 15(0) & 200 & M19,C19 & M19,C19 & DECam,M17 \\
NGC628 & 9.77 & 13(1) & 300 & DECaLS & Gemini & DECaLS \\
NGC3115 & 10.2 & 14(4) & 300 & DECaLS & Gemini & DECaLS \\
NGC5236 & 4.7 & 11(0) & 300 & M15 & M18/DECam & DECam \\
\enddata
\tablecomments{List of the LV hosts considered in this work. The columns list the hosts, distances, and number of satellites (confirmed and unconfirmed) in the mass range $5.5 < \log(M_\star) < 8.5$ and with $M_V<-9$ mag and $\mu_{0,V}<26.5$ mag arcsec$^{-2}$. The imaging data used for dwarf detection, candidate confirmation, and photometry is listed in the final three columns.  The sources listed in these columns are: S18-\citet{smercina2018}, B19-\citet{bennet2019}, M12-\citet{mcconnachie2012}, C13-\citet{chiboucas2013}, M15-\citet{muller2015}, M17-\citet{muller2017}, M18-\citet{muller2018_trgb}, M19-\citet{muller2019}, C19-\citet{crnojevic2019}}
\end{deluxetable*}


\subsection{Auxiliary Galaxy Samples}
\label{sec:aux_data}
To compare with the LV satellite systems, we consider dwarf samples in three auxiliary data sets. First are isolated field dwarfs in the Nearby Galaxy Catalog of \citet{karachentsev2013}. The Nearby Galaxy Catalog includes many satellites of massive LV hosts (including several of the satellites included in the ELVES sample), but here we use it for a sample of isolated field dwarfs. To select a subsample of field dwarfs, we consider only galaxies: 1) with TRGB distances available, 2) with $M_B>-16$, 3) greater than a 3D distance of 500 kpc away from any massive host with $M_B<-18.8$\footnote{This cut was chosen such that all of the ELVES hosts would be included.}, 4) late-type morphology\footnote{which includes the vast majority of these isolated dwarfs.} (see \ref{sec:morph} for definition), and 5) have coverage in either DECaLS or archival CFHT/Megacam images or are included in the catalog of \citet{mcconnachie2012}. In calculating the separation from nearby hosts, we use the minimum 3D distance between dwarf and host for any dwarf distance between 0.9 and $1.1\times$ the reported TRGB distance, in order to account for uncertainty in the TRGB distance. For each of the 121 dwarfs in DECaLS or CFHT/Megacam imaging, we procure cutouts of the dwarfs in the same manner as for the ELVES dwarfs and measure the photometry in the same way (see below). The photometry we use for this isolated dwarf sample is given in Appendix \ref{app:field}. 

The other two data sets we compare with are samples of dwarfs in the nearby Virgo and Fornax clusters. Within the last few years, catalogs of dwarfs with similar completeness limits as ELVES have become available in these clusters, facilitating direct comparison across environments. For Virgo, we use the NGVS galaxy catalog from \citet{ferrarese2020} which considers only the core 4 deg$^2$ region of Virgo (roughly out to $r\lesssim R_\mathrm{vir}/5$). For Fornax, we consider the galaxy catalog of \citet{eigenthaler2018} which considers the inner $r\lesssim R_\mathrm{vir}/4$ region of the cluster.

Both sources provide photometry for the galaxies, however, to maximize the comparability of results in these environments to the ELVES sample, we perform our own photometric measurements. For the NGVS dwarfs, we acquire the raw CFHT/Megacam data from the Canadian Astronomy Data Center\footnote{\url{http://www.cadc-ccda.hia-iha.nrc-cnrc.gc.ca/}}, and reduce the data in the same way we reduce the other CFHT/Megacam data used here \citep[see ][for details on the sky subtraction, etc.]{sbf_calib,carlsten2020a}. Since our focus is primarily on low-mass dwarf galaxies, we only reduce data for the NGVS dwarfs with $M_\star\leq10^9$\msun. In particular, we take the list of dwarf galaxies and stellar masses from \citet{rsj2019}. For the NGFS dwarfs, we use Dark Energy Survey data (specifically the DECaLS reduction) for cutouts of the dwarfs. These DECam data are significantly shallower than the DECam data used by the NGFS team, and thus we only consider dwarfs brighter than $M_g<-10$ for this comparison sample.  We compare our photometry with that of the NGV and NGF Surveys in Appendix \ref{app:ng_compare}. Overall the agreement is excellent with biases in measured sizes $\lesssim1\%$ or so. We do note a bias of $\sim0.3$ mag between our measurements of the Fornax dwarfs and the results of the NGFS. We discuss this more in the Appendix. For both cluster samples, we only consider early type dwarfs, as selected via visual inspection (described below). This removes $<5\%$ of dwarfs for each cluster sample as the populations are overwhelmingly early-type.

\subsection{Completeness of the Surveys}
\label{sec:completeness}
As mentioned above, it is crucial to have well-quantified completeness limits for all of the satellite systems. As discussed in \citet{carlsten2020b}, the common limit of the six hosts surveyed in that work along with the six previously surveyed hosts from the literature is $M_V\sim-9$ mag in luminosity and $\mu_{0,V}\sim26.5$ mag arcsec$^{-2}$ in surface brightness\footnote{Note that this corresponds to an average surface brightness within the effective radius of $\sim27.5$ mag arcsec$^{-2}$ for an exponential profile.}, at a 90\% completeness level. This luminosity limit corresponds to a stellar mass of $\sim10^{5.6}$\msun~ for a mass-to-light ratio of $M_\star/L_V\sim1.2$. For the six hosts surveyed in that work, this was demonstrated with extensive simulations of injecting and recovering mock galaxies \citep[see e.g.][]{carlsten2020a}. The surveys of the very nearby hosts (e.g. MW, M31, CenA, M81) can go significantly fainter and lower in surface brightness since dwarfs can be found via resolved stars but we limit these surveys to this `lowest common denominator' completeness limit. 

We have performed extensive image simulations with the new hosts included in this work, including those surveyed with DECaLS, and have found that a similar limit in both luminosity and surface brightness is appropriate. The exception are the hosts in the BASS portion of DECaLS. BASS is shallower than the other portions of DECaLS and a surface brightness limit of $\mu_{0,V}\sim26$ mag arcsec$^{-2}$ is more appropriate. However, since only 2 of the 24 hosts in the present work use BASS imaging, we consider the $\mu_{0,V}\sim26.5$ mag arcsec$^{-2}$ limit to be representative of the surveys as a whole. 

For several of the hosts surveyed with the CFHT/MegaCam data or from the literature (particularly NGC 4631, NGC 4258, M101, and M94) we have also searched for candidate satellites with the shallower DECaLS data. All of the previously known satellites in these systems above the fiducial completeness limit are readily recovered in the DECaLS search, confirming this completeness limit as robust. 

Note that for all the analysis in this paper, we impose a luminosity cut of $M_V<-9$ and surface brightness cut of $\mu_{0,V}<26.5$ mag arcsec$^{-2}$. In the ELVES survey, we have recovered a number of satellites of fainter luminosity and/or fainter surface brightness, below the nominal completeness limit. However, we do not include those satellites in the analysis undertaken here. 

As mentioned above, while the catalogs of candidate satellites are complete to this level, we do not achieve confident distances for all galaxies within these luminosity and surface brightness limits, and there are a few candidates without meaningful distance information. These are generally some of the faintest candidates for each of the hosts. Throughout this work, we are careful to check that all conclusions are robust to whether these unconfirmed candidates are actually satellites or not. For the results shown in this paper, these unconfirmed satellites are not included.

Regarding the NGVS sample, \citet{ferrarese2020} use artificial galaxy injection simulations to estimate the galaxy catalog is at least $50$\% complete down to magnitudes of $M_g\sim-9$ and surface brightness of $\mu_{0,g}\sim27$ mag arcsec$^{-2}$, comparable to the limits of ELVES. For NGFS, \citet{eigenthaler2018} detected dwarfs visually, precluding a robust estimate of completeness, but given that they detect dwarfs down to $M_g\sim-8$ and have similar point-source depth as NGVS, we assume that the completeness limit is similar. Moving forward, we assume that each survey is roughly complete to the limit of ELVES ($M_V\sim-9$ and $\mu_{0,V}\sim26.5$ mag arcsec$^{-2}$) and take that as the fiducial completeness limit throughout.

As noted in the Introduction, the completeness of the Nearby Galaxy Catalog of \citet{karachentsev2013} is not well-established, and thus any differences between the sample of isolated field galaxies and the late-type ELVES satellites will likely be attributable to incompleteness. However, this sample is still quite useful as a reference, and we include it in the comparisons.

\subsection{Photometry}
\label{sec:photometry}
In this section, we detail how we derive the photometry and structural parameters for the dwarf galaxies considered in this work. In all cases, we rely on parametric 2D S\'{e}rsic profile fits to the surface brightness profiles. While S\'{e}rsic profiles are inadequate for many of the brighter, and particularly star-forming, satellites, the faintness of the majority of the satellites necessitates the use of parametric fits. For consistency, we therefore use S\'{e}rsic profile fits for all dwarfs in the current work. Due to the inadequacy of the S\'{e}rsics for the brightest galaxies, we focus mostly on dwarfs in the mass range $5.5 < \log(M_\star/{\rm M}_\odot) < 8.5$ in this work. 

In all cases, we have two bands for each of the dwarfs, either $g$ and $r$ or $g$ and $i$. We fit the S\'{e}rsic profiles in the manner of \citet{sbf_calib, carlsten2020a} using \textsc{imfit} \citep{imfit}. The $g$-band image is masked for nearby stars and background galaxies and fit with a S\'{e}rsic profile. This initial masking also includes point sources that are likely galactic nuclei. Many of the dwarfs in the sample are nucleated, and we investigate the properties and prevalence of nuclei in a companion paper (Carlsten et al., submitted). For dwarfs that do exhibit a central nucleus (defined as having a point source within $\lesssim r_e/8$ of the photometric center of the galaxy), we refit the galaxy with two S\'{e}rsic profiles where the second profile (representing the nucleus) is restricted to have $r_e\lesssim1$\arcs. For nucleated dwarfs, the photometry we consider in this paper comes from the first S\'{e}rsic, representing the diffuse stellar body of the galaxy. Thus the central surface brightness that we quote for nucleated galaxies is the central surface brightness resulting from extrapolating the S\'{e}rsic profile to the center (i.e. not including the nucleus). The $r$ or $i$-band image is then fit, allowing only the amplitude to change.

In our experience, the initial masking threshold and image cutout size used in the fit can have a sizable impact on the inferred S\'{e}rsic parameters. Thus, we use the following prescription to set these for each dwarf in a determinate way. Initial guesses are set visually for each dwarf to get a stable, acceptable fit. Then, we iteratively refit using a cutout size of $3r_e \times 3r_e$, where $r_e$ is the effective radius of the galaxy model. In each iteration, we subtract out the current best-fit galaxy model and mask using a threshold of $0.5\times$ the peak surface brightness of the galaxy model. The galaxy model is added back in and the galaxy is refit with this new mask and cutout size. We continue these iterations until the effective radius changes by less than $5$\% between iterations. Most galaxies are fit within 3 iterations with the effective radius changing by $\lesssim15\%$, in total.

The photometry is all corrected for extinction using the maps of \citet{sfd2} and is in the AB system. The effective radii that we present are all measured along the major axis.

To deal with the fact that sometimes we have $g/r$ and sometimes we have $g/i$, we make use of the conversion 
\beq
g-i = 1.53 (g-r) - 0.032
\eeq
which we derive from MIST \citep{mist_models} SSP models for a range in age between 3 and 10 Gyr and metallicities in the range $-2<\mathrm{[Fe/H]}<0$\footnote{This is calculated specifically for the CFHT filter system.}. Due to the relative uncertainty of this conversion we do not heavily rely on the colors of the dwarfs in our analysis. Due to historical consistency, we generally report our photometry in $V$-band for which we use the transformation for the SDSS filters\footnote{\url{http://www.sdss3.org/dr8/algorithms/sdssUBVRITransform.php\#Lupton2005}}:
\beq
V = g - 0.5784(g-r) - 0.0038
\eeq

The specific dataset used for photometry for each host is given in Table \ref{tab:hosts}. Even for hosts that have significantly deeper imaging data (for applying SBF), we generally  use the DECaLS data, if available, to keep the photometry and filter systems as consistent as possible. With that said, when comparing the photometry from DECaLS with the photometry from CFHT for overlapping dwarfs where we have S\'{e}rsic fits from both datasets (about 70 dwarfs in common, including dwarfs that the SBF constrains to be background), we find good agreement. The measurements of luminosity and size are unbiased between the two sources and have scatter similar to the estimated uncertainty in the measurements. 

Another concern is that we are dealing with several different filter systems (DECam, BASS, MegaCam). All are Sloan-like filters but will have some differences. We make no attempt to bring the different measurements onto the same filter system, instead opting to show in Appendix \ref{app:filters} that the differences in the filter systems do not alter our conclusions. As said above, we do not heavily rely on colors specifically because colors will be the most significantly affected by filter system differences due to the much smaller dynamic range in color than luminosity. We give the photometry in tables in Appendix \ref{app:photo_tables}, carefully accounting for the sources of each measurement so that one could make the filter transformations later, if necessary.

For a small subset of dwarfs, a bright foreground star or other contaminant precluded a robust S\'{e}rsic fit. We note in the photometry tables in the appendix which dwarfs are particularly suspect. We still include them in most of the analysis. However, this does not change the results of the paper due to the small number of affected galaxies. 

\subsection{Uncertainty in Structural Parameters}
\label{sec:uncertainty}
To derive robust scaling relations between structural parameters and, particularly, to probe the intrinsic scatter present in scaling relations, it is critical to have accurate estimates of the uncertainties in the measured structural parameters. Estimating uncertainties in the photometry of very low luminosity and low surface brightness dwarfs is notoriously difficult as the measurements are generally entirely systematics dominated. The quality of the sky subtraction and the presence of nearby contaminating sources significantly affect the photometric measurements. 

Arguably the most robust way to estimate uncertainties is to inject artificial galaxies and measure how well the input parameters are recovered. Ideally the artificial galaxies are injected prior to sky subtraction to be able to quantify the effect of imperfect sky subtraction on the dwarf photometry. This was the approach for the dwarfs studied in \citet{carlsten2020a, carlsten2020b}, and here we use the uncertainties reported in that work for the relevant hosts. There, for each dwarf we first measured the photometry from the S\'{e}rsic fit and then injected artificial galaxies with the same S\'{e}rsic parameters into the raw CCD-level data (i.e. before sky subtraction and coaddition). These artificial galaxies were then fit with S\'{e}rsic profiles, and the uncertainty in the original S\'{e}rsic parameters were estimated by the spread in the recovered parameters.  

For the hosts not previously considered in \citet{carlsten2020a, carlsten2020b}, due to the prohibitively large number of dwarf satellites, we do not follow the same, dwarf-by-dwarf procedure. Making use of the relatively consistent depth of imaging across the DECaLS footprint (which most of these hosts use for photometry, cf. Table \ref{tab:hosts}), we do one set of simulations, injecting dwarf galaxies  and quantifying the spread in the recovered photometry as a function of apparent magnitude. One tract (3600$\times$3600 pixel images which form the basis of DECaLS) is randomly chosen from the survey footprint of each of the DECaLS hosts. Using these tracts as the background images, we inject artificial dwarf galaxies that roughly follow the stellar mass-effective radius relation (including scatter) that we find in this work. Colors, ellipticities, and S\'{e}rsic indices are also drawn from distributions roughly matching what we find in this work for the real dwarfs. The spread in recovered parameters is quantified as a function of the apparent magnitude of the input dwarf. These functions are shown in Appendix \ref{app:uncertainty} and used for the real dwarfs to derive uncertainties. Since these dwarfs are injected after coaddition and sky-subtraction, the estimated uncertainty gets very close to zero for the brightest dwarfs. To deal with this, we use the estimated uncertainties from the more realistic image simulations of \citet{carlsten2020a, carlsten2020b} at the bright end, effectively introducing a ``floor" in the uncertainty due to sky-subtraction related problems. More details of this process are given in Appendix \ref{app:uncertainty}. Note that for the couple other hosts which were not in the \citet{carlsten2020a, carlsten2020b} sample nor covered in DECaLS (e.g. NGC 6744 and CenA), we use these same functions to estimate uncertainties. In most cases, the non-DECaLS data are deeper, and thus the uncertainties are conservative estimates.

\subsection{Stellar Masses}
\label{sec:mstars}
We derive stellar masses of the dwarf satellites from integrated luminosity and color using a color-$M_\star/L$ relation. In particular, we use the CMLRs derived by \citet{into2013} for a stellar population with an exponential SFH. For the two filter combinations we use in this work, these are:
\beq
\label{eq:cmlr}
\log(M_\star/L_g) = 1.774 (g-r) - 0.783\\
\log(M_\star/L_g) = 1.297 (g-i) - 0.855\\
\eeq
We take the solar absolute magnitude in $g$ to be 5.03 \citep{willmer2018}\footnote{This is specifically the value for the CFHT $g$-band, but we use it for all the dwarfs in this work, regardless of telescope.}. We propagate the uncertainties in both the luminosity and color of each dwarf to estimate uncertainty in the stellar mass.

We use the CMLRs from \citet{into2013} for all stellar masses compared in this work, including those of dwarfs from outside of the ELVES survey, so we do not expect that the details of the CMLR (e.g. what SFH is used in the model) will have a significant impact on the analysis. With that said, in Appendix \ref{app:ng_compare}, we compare the stellar masses derived via Equation \ref{eq:cmlr} for NGVS and NGFS dwarfs and those derived by those respective collaborations using different CLMRs (and generally more than just two bands). We find disagreements of the order $\sim0.1$ dex (likely due to different IMF choices). Thus we include 0.1 dex as an additional systematic uncertainty in the stellar mass measurements.

For dwarfs without color information, we use relations between $M_\star/L_V$ and $M_V$ that we separately determine from the early-type and late-type dwarfs (see next section) that do have color information \citep[and using the CMLRs from][]{into2013}. In particular, we find

\beq
\left( M_\star/L_V\right)_\mathrm{etg} = -0.096 M_V + 0.229 \\
\left( M_\star/L_V\right)_\mathrm{ltg} = -0.083 M_V - 0.4528 \\
\eeq
These ratios broadly agree with what has been used for LG dwarf satellites in this mass range before \citep[e.g.][]{woo2008}. For galaxies without color, we assume a constant uncertainty of 0.2 dex in the stellar mass.

\begin{figure*}
\includegraphics[width=\textwidth]{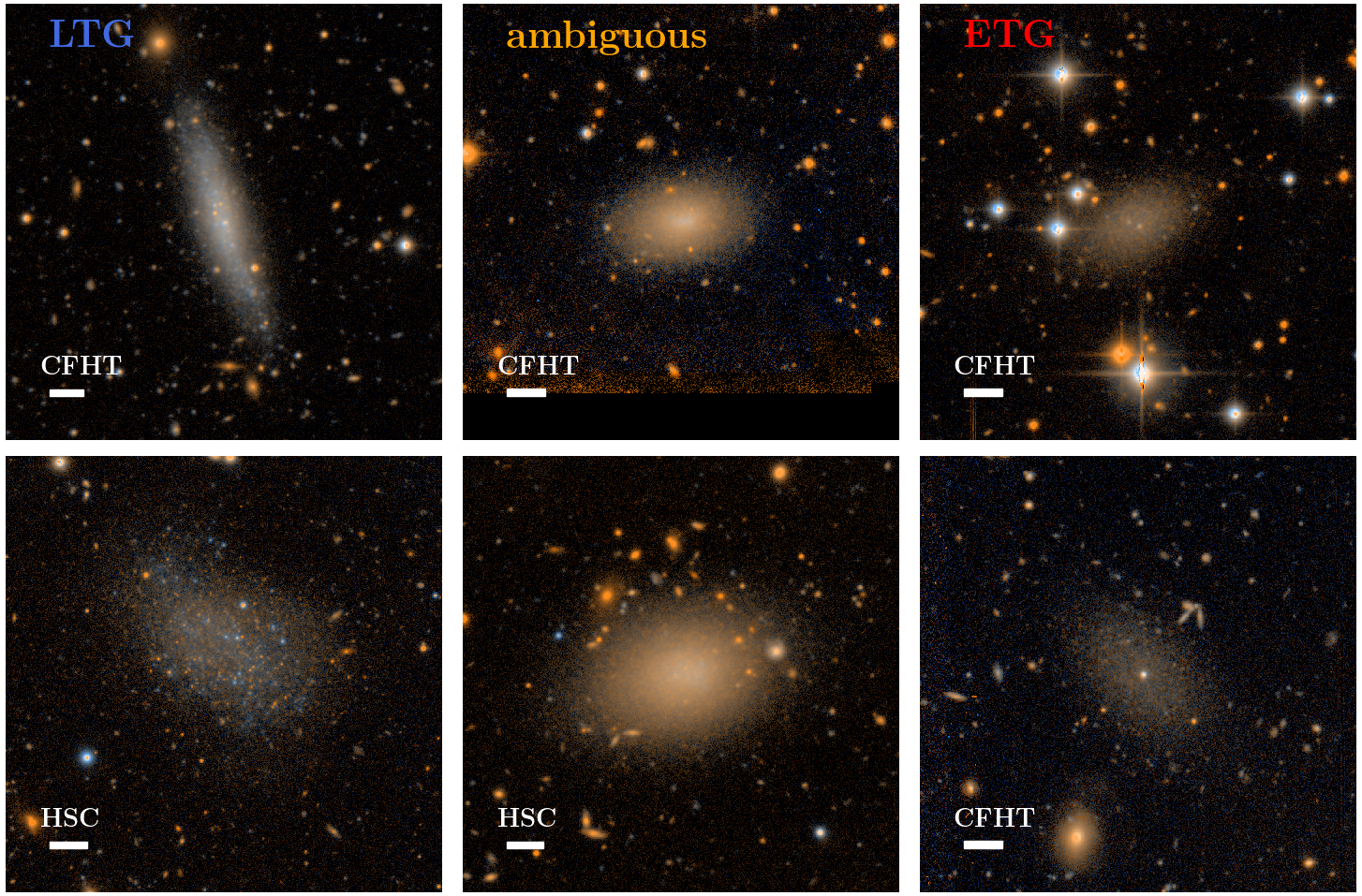}
\caption{Examples of dwarfs visually classified as `early-type' and `late-type'. Late-type dwarfs are irregular, with apparent active star formation throughout the galaxy while early-types exhibit smooth surface brightness profiles without any clear blue, star-forming clumps. The middle column (`ambiguous') shows two dwarfs that eluded a simple classification. These two overall possessed the smooth round surface brightness profile of early-type dwarfs but had a noticeable bluish tint (and/or a kink in the brightness profile) in the central region of the galaxy, possible indicating recent or ongoing star formation. For these galaxies, we relied instead on the color-magnitude relations of late-types and early-types to classify them (cf. Appendix \ref{app:morph}). The white bar in each panel denotes $10$\arcs. }
\label{fig:examples}
\end{figure*}

\subsection{Galaxy Morphology}
\label{sec:morph}
The satellites considered in this work constitute both a late-type, star forming population and an early-type, quenched population (LTGs and ETGs, henceforth). Because it is expected that these generally form an evolutionary sequence (LTGs fall into their hosts, are quenched, and turn into ETGs), it is of significant interest to separately consider the structures and scaling relations of these two sub-populations separately. 

We split the dwarfs into these two groups based on a visual inspection of the dwarf morphology. Given the generally quite deep imaging data available (deep enough to apply SBF) and proximity of these dwarfs, we believe this is the most robust way to split dwarfs with the data available. Dwarfs with smooth, feature-less morphology are classed as early-types while dwarfs with clear star-forming regions, blue clumps, dust-lanes, or any other kinks in their surface brightness profile are classed as late-type. While we generally use the DECaLS data for the photometry, we use all the imaging available (including the deeper imaging used for SBF) for this visual classification. Color images (either $g/r$ or $g/i$ composites) of example dwarfs classified as early- and late-type are given in Figure \ref{fig:examples}. Note that both ETGs and LTGs appear amenable to single S\'{e}rsic profile fits.

These two classes also separate clearly in color-magnitude space. The color magnitude relation, $g-i=-0.067\times M_V - 0.23$, splits the dwarfs into two classes with the same result as the visual inspection about 90\% of the time. 

In most cases the distinction between the two classes is unambiguous. However, there is a subpopulation (roughly $5-10$\% of the whole satellite sample) of dwarfs that would best be classified as ``transition'' objects. These objects generally are round and smooth like the early-type dwarfs but might have what appear to be star-forming regions identifiable as blue clumps in the color images. To be as consistent as possible, for these we simply rely on the color-magnitude division given in the preceding paragraph. Two of these ambiguous cases are shown in Figure \ref{fig:examples}. We show the color-magnitude diagram and dividing line in Appendix \ref{app:morph}.

The presence of H$\alpha$ emission or a large reservoir of neutral gas would be a more robust indicator of star formation activity \citep[e.g.][]{grcevich2009, geha2017, mao2020, karunakaran2020}. However, since our survey is primarily photometric-based at this point, many of the dwarfs do not have spectra or measurements of their neutral gas. About a quarter of the satellites have archival spectra (most often through SDSS) and about a third have useful H\textsc{I} measurements. We will present a more detailed analysis of these measurements in a future work, but suffice to say that they closely corroborate the visual morphology classification. We find that $\sim95$\% of the early-type dwarfs with spectra do not have significant H$\alpha$ emission (equivalent widths $<2$\AA) while $\sim85$\% of the late-type dwarfs with spectra do have H$\alpha$ above this level\footnote{Note that more might have H$\alpha$ emission that is simply below the sensitivity of the archival spectra (generally from SDSS).}. Almost all of the early-type dwarfs that have neutral gas measurements that are deep enough to be sensitive to gas reservoirs of $M_\mathrm{HI} \sim 2\times M_\star$ have neutral gas fraction upper limits below $M_\mathrm{HI}/M_\star\lesssim0.5$.

\section{Dwarf Structural Results}
\label{sec:results}
In this section, we present the photometry and structural parameters of the dwarf galaxy samples. We start by showing 1D surface brightness (SB) profiles of dwarfs in different stellar mass ranges to give an overview of the data and show the salient differences between the early-type and late-type galaxies. Second we compare the scaling relations between size, SB, S\'{e}rsic index, luminosity, and stellar mass for early-type versus late-type satellites at fixed environment. We then compare the scaling relations across different environments, fixing galaxy type. Third, we show the distribution of sizes at fixed stellar mass. Finally, we provide power-law fits to the various mass-size relations presented.

\begin{figure*}
\includegraphics[width=0.98\textwidth]{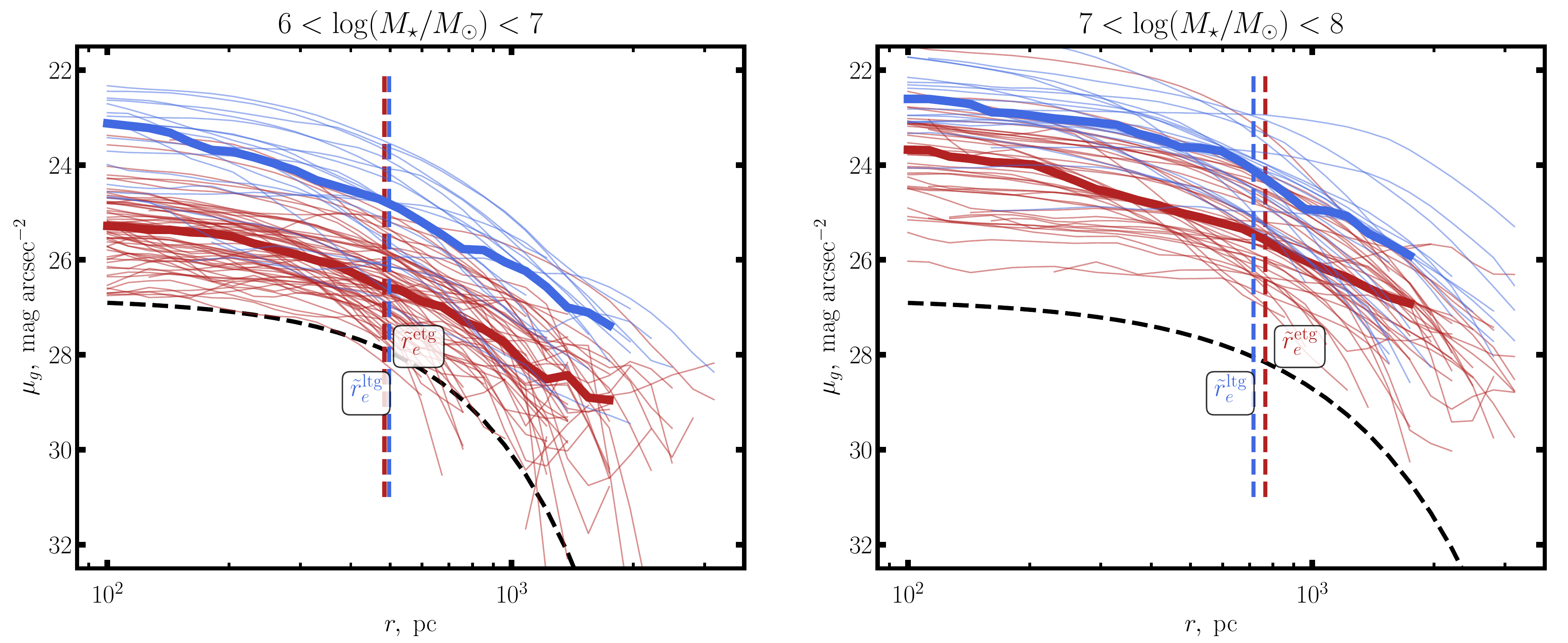}
\caption{The 1D surface brightness profiles for dwarfs in two mass bins split into late-type (blue) and early-type (red) dwarfs. The thick lines in each panel show the running median surface brightness profile. In both mass bins, the late-type dwarfs are $\sim2$ mag brighter in surface brightness than the early-type dwarfs but have very similar median effective radii, indicated by the vertical dashed lines. The black dashed lines roughly indicate the surface brightness limit of the survey, shown as a S\'{e}rsic profile with $\mu_{0,V}=26.5$ mag arcsec$^{-2}$ and the median $n$ and $r_e$ of the observed early type galaxies.}
\label{fig:1d_profiles}
\end{figure*}

\subsection{1D Surface Brightness Profiles}
\label{sec:sb-profiles}
Figure \ref{fig:1d_profiles} shows the average one dimensional surface brightness profiles for ELVES dwarfs split into two stellar mass bins. We extract the $g$-band 1D SB profiles with elliptical annuli using the ellipse parameters from the S\'{e}rsic fits. Profiles are extracted for each galaxy within the radial range $0.1r_e<r<3r_e$ but are interpolated onto a common grid in physical radius in Figure \ref{fig:1d_profiles}. The few dwarf galaxies with significant interference from nearby stars (these are indicated in the photometry tables in Appendix \ref{app:photo_tables}) are not included in Figure \ref{fig:1d_profiles}. Due to the varying distances to the dwarfs and varying data quality, we do not indicate the spatial scale of the PSF for comparison. At the average distance of $\sim7$ Mpc, the inner radius of 100 pc is $\sim3$\arcs~ which is significantly bigger than the typical PSF size ($\lesssim1$\arcs). Note that the inner radius of 100 pc precludes any central nucleus in the SB profile. 

The late-type dwarfs are significantly higher surface brightness than the early-type, by generally $\sim2$ mag, but are roughly the same size in terms of effective radius in these mass bins. This result that late-type and early-type dwarfs are the same size at fixed stellar masses is further explored in the following sections and is one of the main results of this work. Additionally, in later sections, we show that the $\sim2$ mag difference in SB can be explained simply by passive aging of the stellar population. This idea that the late-type dwarfs can lead to the early-type dwarfs without much (or even any) structural changes beyond simply quenching is a recurring theme in this work.

\begin{figure*}
\includegraphics[width=0.98\textwidth]{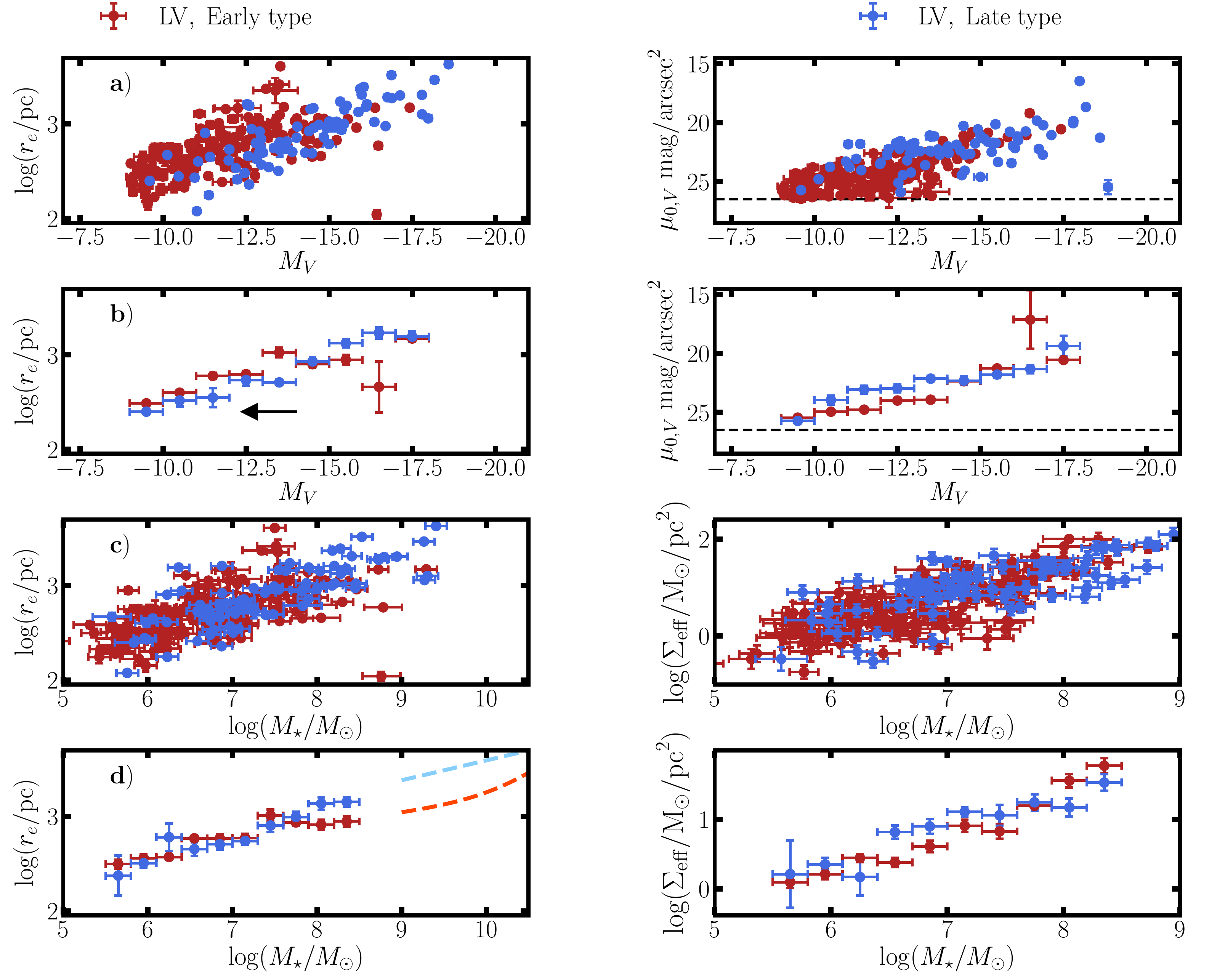}
\caption{\textit{Rows a \& b}: Scaling relations of effective radius and surface brightness with luminosity for the LV dwarf samples, split into late-type and early-type (blue and red, respectively). Row \textit{b} shows average trends binned in 1 mag wide bins of luminosity. In this row, the vertical errorbars indicate the error in the mean, not the intrinsic spread. The arrow in the left panel shows the 1.4 mag change due to passive aging of a [Fe/H]$=-1$ stellar population from 1 Gyr to 6 Gyr. The SB limit of the LV satellite sample of $\mu_{0,V}<26.5$ mag arcsec$^{-2}$ is shown in the right panels. \textit{Rows c \& d}: : Scaling relations of effective radius and effective stellar surface density ($\Sigma_{\rm eff}\equiv M_\star/2\pi r_e^2(1-\epsilon)$) with stellar mass. Row \textit{d} shows average trends binned in 0.3 dex wide bins of stellar mass.  In the left panel, the dashed lines show the mass-size relations for early-type (red) and late-type (blue) galaxies from the GAMA Survey \citep{lange2015}.}
\label{fig:etg_v_ltg_lv}
\end{figure*}

\subsection{Late-type vs. Early-type Structure}
\label{sec:etg_v_ltg}
In this section, we compare the structure of late-type dwarfs to early-type at fixed environment, namely in LV satellite systems. In Figure \ref{fig:etg_v_ltg_lv}, we show the scaling relations of size and surface brightness both as functions of integrated luminosity and stellar mass (top panels and bottom panels, respectively). We show the individual ELVES dwarf satellites along with the average trends in bins of luminosity or stellar mass. Both early-type and late-type satellites show clear and strong trends between both effective radius and central surface brightness with luminosity and stellar mass. We emphasize several salient features in this figure.

First, the fraction of late-type dwarfs is significantly higher at larger stellar masses and, especially, higher luminosities. Above, we argued that the division into late-type and early-type morphology is essentially a physical distinction between actively star-forming and quenched dwarfs. Thus, the quenched fraction of dwarfs is clearly a strong function of dwarf luminosity. This trend can also be seen versus stellar mass, so it is not simply on effect of differing mass-to-light ratios. We interpret this trend to mean that the quenching timescales for lower luminosity (hence, lower mass) dwarfs is shorter than for higher mass dwarfs, as has been argued from observations of the Local Group and hydrodynamic simulations \citep[e.g.][]{wetzel2015,fillingham2015,akins2020}. We relegate a further discussion of this to a future paper that will also present measurements of the gas properties of the dwarf satellites.

Second, the distribution of dwarfs in surface brightness clearly runs into the SB limit of $\mu_{0,V}<26.5$ mag arcsec$^{-2}$ as shown in the upper right panel. The average surface brightness appears to reach this limit around $M_V\sim-9$, but due to the scatter around the average relation, ELVES is likely missing dwarfs due to SB incompleteness at brighter magnitudes, perhaps even at $M_V\sim-12$. We will address this SB incompleteness further in \S\ref{sec:sizes} where we consider the distribution of sizes at fixed stellar mass.

Third, at the faint end, the late-type satellites are generally smaller at fixed luminosity than the early-type dwarfs. This is a reversal of the known behavior for more massive dwarfs ($M_\star\sim10^9$\msun) where late-type dwarfs are larger than early types \citep{shen2003, lange2015}. This reversal is somewhat visible around $M_V\sim-15$, although the dearth of early-type dwarfs above this luminosity (which includes M32, the extreme outlier in size in the top left panel) precludes a firm conclusion of where the change in size begins. Low-luminosity LTGs were seen to be smaller than ETGs in the Fornax Deep Survey as well \citep{venhola2019}. In the bottom right panel, the arrow shows the luminosity change due to passive aging of a [Fe/H]$=-1$ stellar population from 1 Gyr to 6 Gyr as predicted by the MIST isochrone models \citep{mist_models}. These parameters are chosen to roughly match the average $g-i$ colors of the ETGs and LTGs ($\sim0.8$ and $\sim0.5$ respectively). The actual average ages of the late-type and early-type groups are not known, but the fact that this passive evolution is roughly the right magnitude to move the late-type dwarfs to the early-type trend indicates that it is likely that the difference in sizes at fixed luminosity is mostly a stellar population effect. 

Indeed, at fixed stellar mass, the late-type and early-type dwarfs show very similar average sizes. There is no clear offset between late-type and early-type dwarfs as there was for the luminosity-size relation, confirming our assertion that the offset was largely an effect of differing mass-to-light ratios.

Also in Figure \ref{fig:etg_v_ltg_lv}, we show the mass-size relations for more massive ETGs and LTGs from the GAMA Survey \citep{lange2015}. A single power law is used for the LTGs while a double power law was found in that work to be more appropriate for the ETGs. The specific LTG/ETG split that we plot in Figure \ref{fig:etg_v_ltg_lv} comes from a cut in morphology \citep[as described in][]{lange2015}, but the qualitative behavior is unchanged if a different metric for classification is used instead. Similar qualitative results are also seen in the mass-size relation for LTGs/ETGs from SDSS \citep{shen2003}. These works show that LTGs are larger than ETGs at fixed stellar mass, at least in the range $10^{9}\lesssim M_\star/M_\odot<10^{11}$ but that the slope of the ETG mass-size relation is shallower at the low-mass end and will presumably intersect with the LTG mass-size relation at some lower dwarf mass. From our results, this intersection is somewhere in the vicinity of  $M_\star\sim10^{8}$\msun. Around this mass range, the slope for the ETGs steepens again, as found also by \citet{eigenthaler2018} for dwarfs in the Fornax cluster. At lower masses, both the LTG and ETG populations show a very similar normalization and slope to the mass-size relation. 

In the bottom right panels of Figure \ref{fig:etg_v_ltg_lv}, we show the (projected) stellar mass density within the effective radius. We calculate this as $\Sigma_{\rm eff}\equiv M_\star/2\pi r_e^2(1-\epsilon)$,  including a term with the ellipticity to account for the fact that the effective radius is taken along the major axis. The late-type sample is generally shifted to higher stellar densities, particularly at intermediate stellar masses. Since the effective radii are similar between the late-type and early-type samples, this offset is largely coming from differences in the average ellipticity. We explore this in more detail in Section \ref{sec:shapes}.

\begin{figure}
\includegraphics[width=0.48\textwidth]{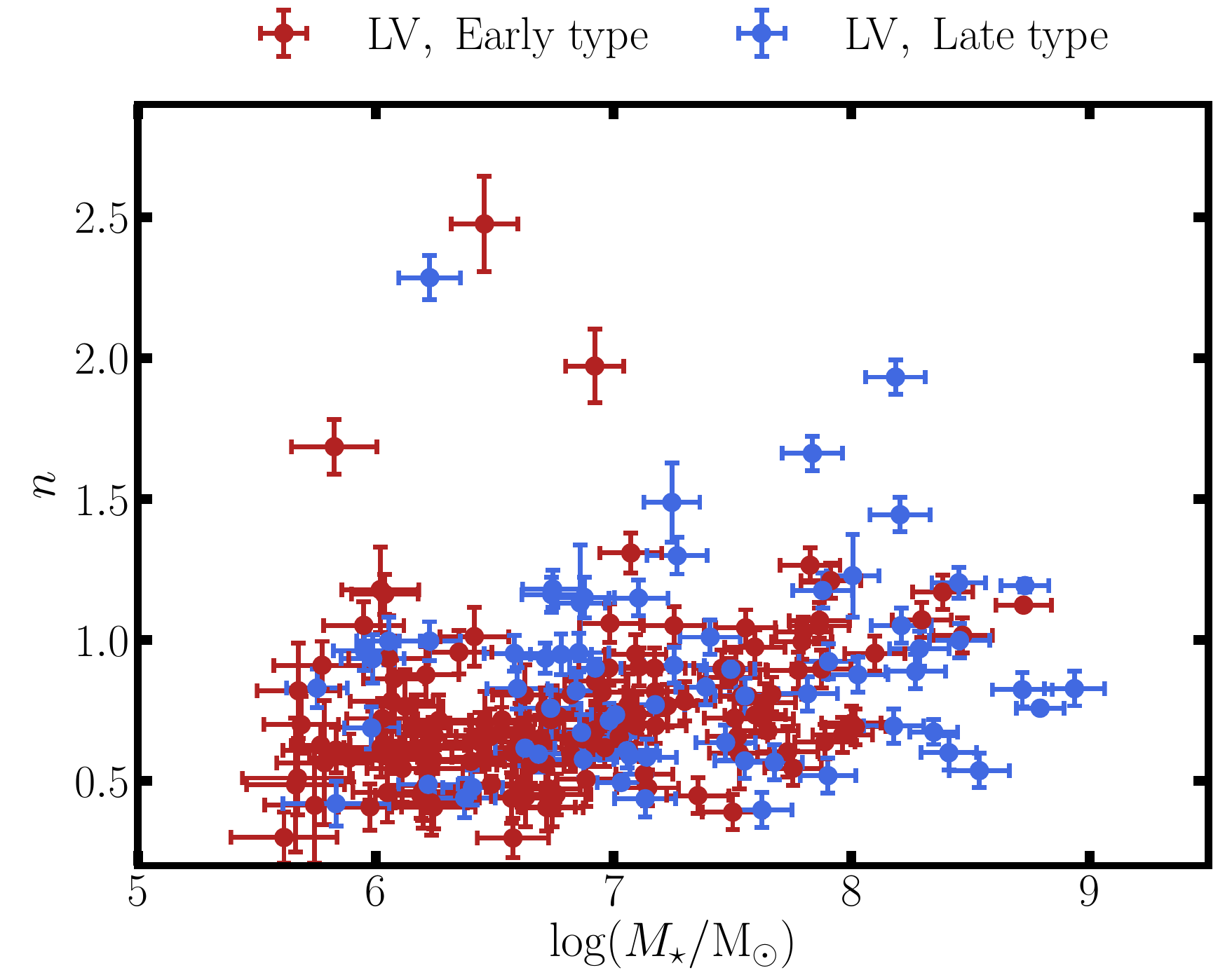}
\caption{S\'{e}rsic indices versus stellar mass for the two LV dwarf samples.  }
\label{fig:sersic-n-lv}
\end{figure}

Figure \ref{fig:sersic-n-lv} shows the distribution of S\'{e}rsic indices for the LV ETG and LTG populations plotted against stellar mass. The early-type and late-type LV satellites overall show quite overlapping distributions in S\'{e}rsic index, although the early-types show a trend of increasing $n$ for larger stellar masses while the late-types show less noticeable of a trend.

\begin{figure*}
\includegraphics[width=0.98\textwidth]{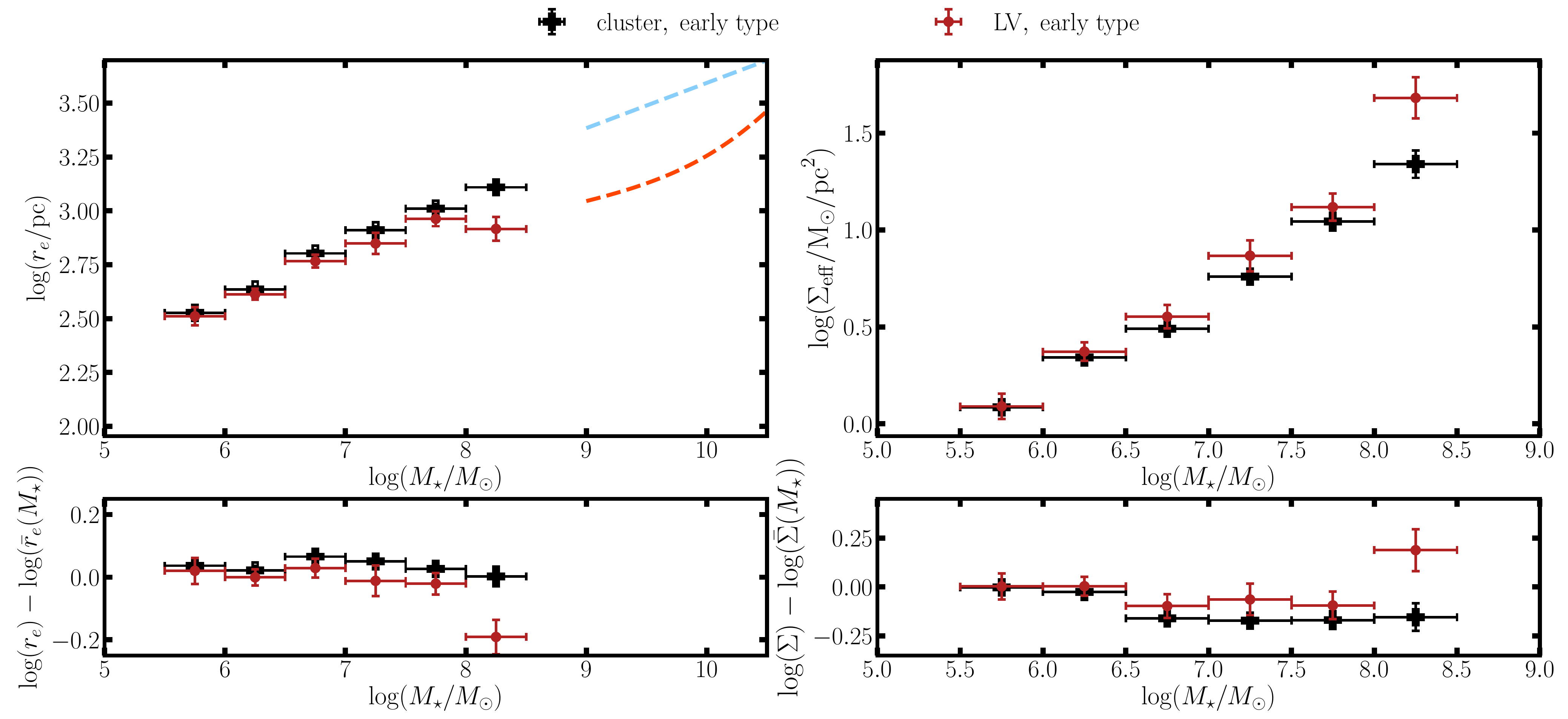}
\caption{Scaling relations of effective radius and effective stellar surface density ($\Sigma_{\rm eff}\equiv M_\star/2\pi r_e^2(1-\epsilon)$) with stellar mass. The LV early-type satellite dwarfs are compared to early-type dwarfs in the joint Virgo and Fornax clusters sample. The data points show average trends binned in 0.5 dex wide bins of stellar mass. In these panels, the vertical errorbars indicate the error in the mean, not the intrinsic spread. In the top left panel, the dashed lines show the mass-size relations for early-type (red) and late-type (blue) galaxies from the GAMA Survey \citep{lange2015}. The bottom panels show the residuals after subtracting out general trends from fitting the entire LV sample.}
\label{fig:cl_v_lv}
\end{figure*}

\subsection{Effect of Environment}
\label{sec:environ}
In this section, we consider the effect of environment on dwarf structure at fixed dwarf type. Thus, we compare both cluster ETGs to LV satellite ETGs and field LTGs to LV satellite LTGs. Since stellar mass is a more fundamental quantity than luminosity, we just consider the scaling relations of size and stellar density with stellar mass.

Before showing the results, it is worth clarifying the differences between the LV and cluster environments. The LV hosts are significantly lower in halo mass than either cluster. While host selection in ELVES is simply based on host $M_K$ (corresponding to $M_\star^{\rm host}\gtrsim1/2M_\star^{\rm MW}$), we expect most of the ELVES hosts are roughly MW-sized in halo mass, $M_{\rm halo}\sim10^{12}$~\msun. There are, however, a number of richer hosts (``small-group'' sized) like the M81/M82 system that are several times more massive with $M_{\rm halo}\sim5\times10^{12}$~\msun~\citep{karachentsev2014_masses}. Likely, the most massive halo in the LV is the Leo I group (for which we take NGC 3379 as the `host') at $M_{\rm halo}\sim10^{13}$~\msun~\citep{kourkchi2017}. On the other hand, the Fornax cluster has a dynamical mass of $M_{200}\sim7\times10^{13}$~\msun~\citep{kourkchi2017}, noticeably less massive than the Virgo cluster which has an estimated mass closer to $5-6\times10^{14}$~\msun~\citep{ferrarese2012, kourkchi2017, kashibadze2020}. Thus, the LV satellites are in halos generally at least an order of magnitude less massive than the environment of the cluster satellites. The environmental difference is likely exacerbated by the fact that the cluster samples are from the very central $r_\mathrm{proj}\lesssim R_\mathrm{vir}/4$ regions while the LV host satellite samples are generally complete to at least $r_\mathrm{proj}\gtrsim R_\mathrm{vir}/2$.

Figure \ref{fig:cl_v_lv} shows the scaling relations of size and stellar surface density with stellar mass for early-type dwarfs in the clusters versus in the LV satellite systems. The average sizes are very similar across environment at fixed stellar mass. However, there is a consistent trend of cluster dwarfs being slightly larger, and lower stellar density, than LV satellite dwarfs. We will quantify this slight offset below in Section \ref{sec:msr}.

Numerous previous studies \citep[e.g.][]{smith-castelli2008, misgeld2008, misgeld2011, eigenthaler2018} have shown that the ETG mass-size relation is shallowest at intermediate masses $10^{8}<M_\star/M_\odot<10^{9.5}$ with relatively constant sizes of ETGs of $\sim1$ kpc over this range in stellar mass. \citet{eigenthaler2018} showed from dwarfs in the Fornax cluster that the ETG mass-size relation steepens significantly at lower masses. This behavior is upheld for the early-type LV satellites as well. Unfortunately, due to the lack of early-type satellites with $M_\star>10^8$\msun~ in MW-sized host halos, it is unclear if the turn-down occurs at the same stellar mass or not.

\begin{figure*}
\includegraphics[width=0.98\textwidth]{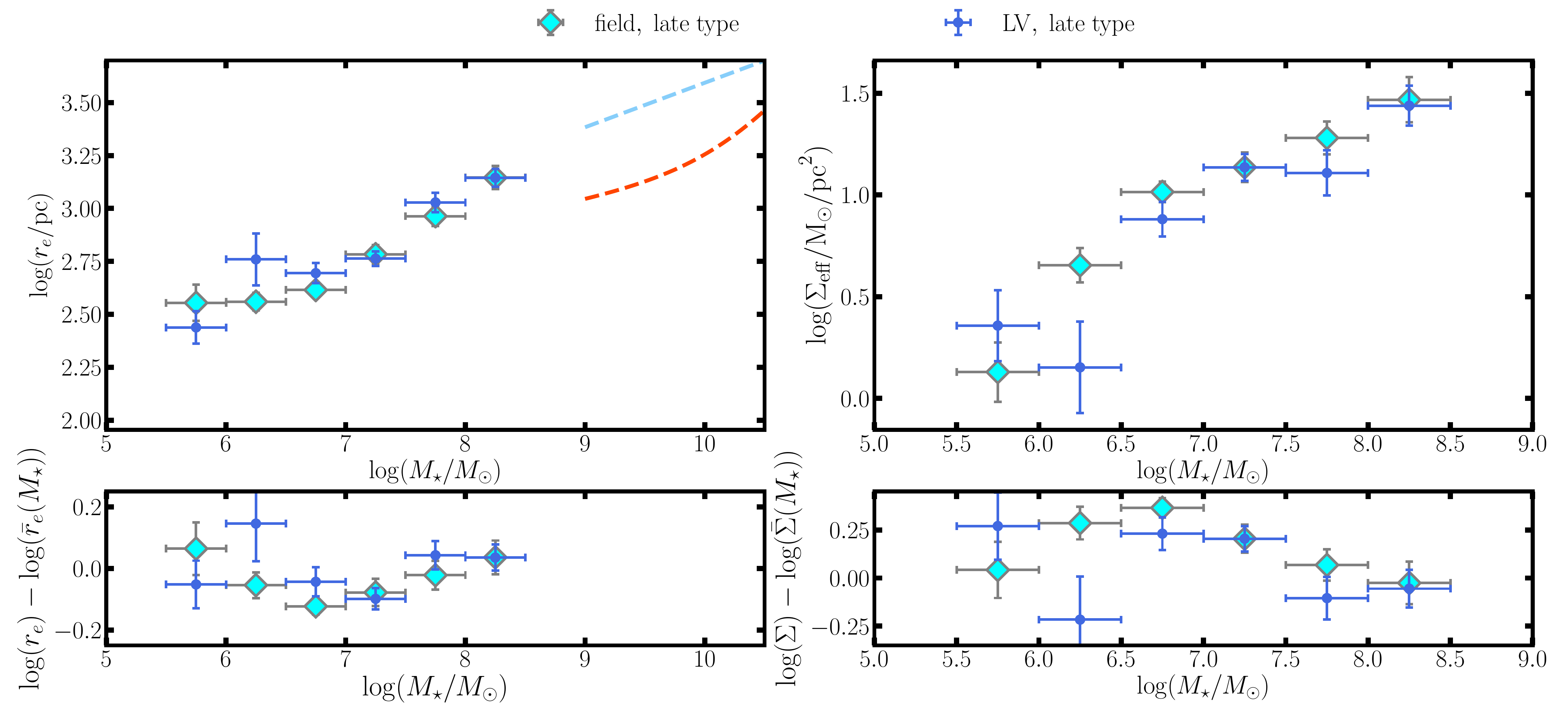}
\caption{Analogous to Figure \ref{fig:cl_v_lv} except now showing the LV late-type satellite dwarfs compared to isolated late-type dwarfs in the field.}
\label{fig:field_v_lv}
\end{figure*}

Figure \ref{fig:field_v_lv} shows an analogous environmental comparison, this time between the isolated field galaxies from the Nearby Galaxy Catalog \citep{karachentsev2013} and the late-type LV satellites. Overall these two environmental samples show very similar scaling relations, within the spread of the observations. Comparing with Figures \ref{fig:etg_v_ltg_lv} and \ref{fig:cl_v_lv}, both samples show higher stellar mass density than the early-type dwarfs, particularly at intermediate stellar masses ($M_\star\sim10^7$\msun). There does not appear to be a consistent offset in size between these two samples.

\begin{figure}
\includegraphics[width=0.48\textwidth]{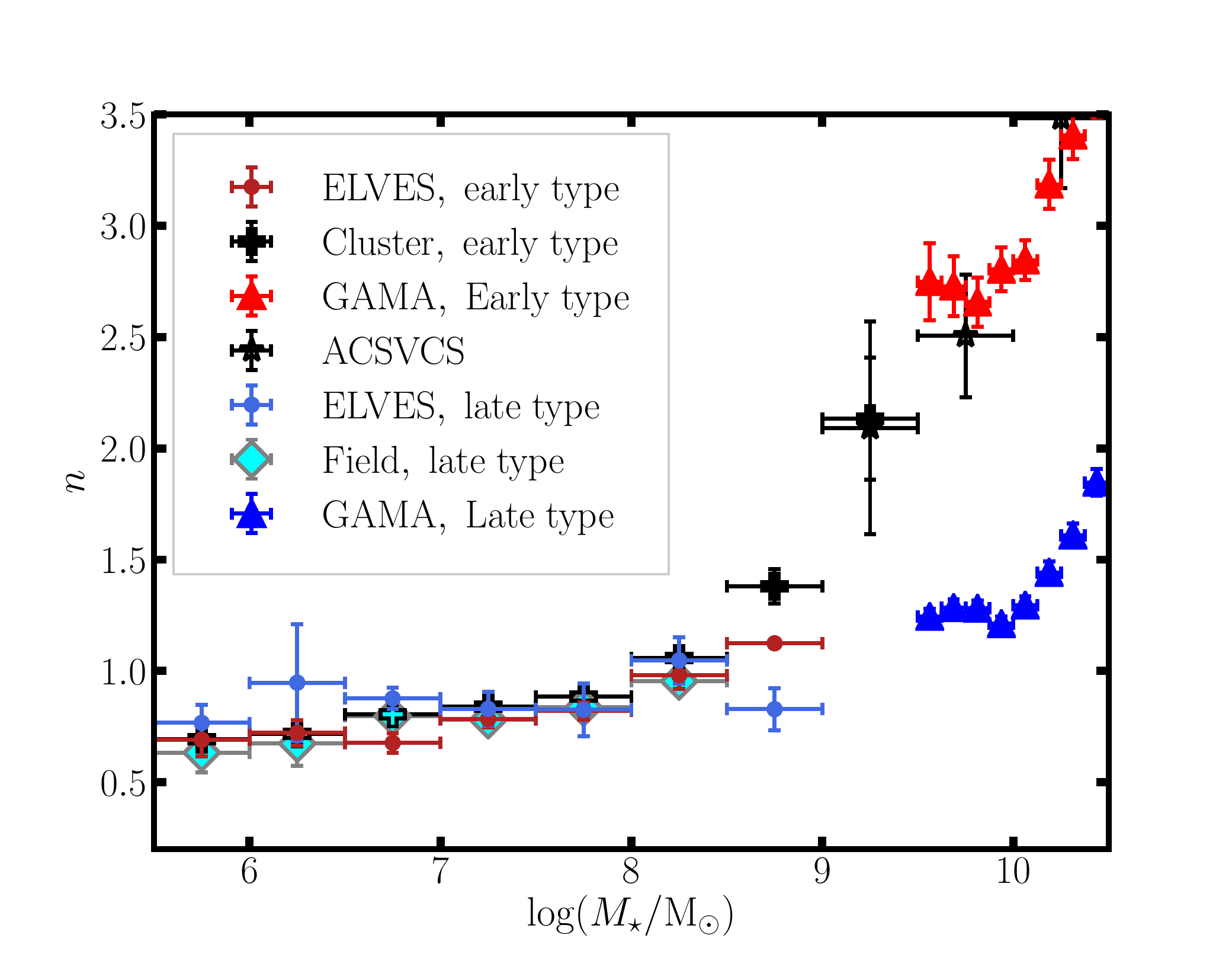}
\caption{S\'{e}rsic indices versus stellar mass for the different dwarf samples considered here.  The average index is shown in 0.5 dex wide bins of stellar mass for each population. The late-type galaxies generally exhibit low indices $n\lesssim1$ while the early-types exhibit increasing values of $n$ for larger stellar masses, particularly above $\sim10^{8.5}$\msun. The ``cluster early type'' points refer to the combined NGVS and NGFS samples.}
\label{fig:sersic_n_all}
\end{figure}

Figure \ref{fig:sersic_n_all} shows the distribution of S\'{e}rsic indices for the different environment samples considered here plotted against dwarf stellar mass, all shown as running averages. We complement the dwarf samples with various samples of higher mass galaxies from the literature. We include the results from the GAMA survey \citep{driver2011} taking S\'{e}rsic indices and stellar masses from \citet{kelvin2012} and \citet{taylor2011}, respectively. The GAMA galaxies are split into early- and late-type based on visual inspection (essentially split as `elliptical' and `not elliptical'). We also include the results from the ACSVCS \emph{HST} Survey of Virgo galaxies from \citet{ferrarese2006}\footnote{Note that there will naturally be some overlap in galaxies between this sample and the NGVS sample.}. 

Interestingly, the different environments show very similar trends for fixed galaxy type. All early-type galaxy samples show a trend of increasing index with stellar mass throughout the mass range, but the trend steepens significantly around $M_\star\sim10^{8.5}$\msun.  At smaller masses ($M_\star\lesssim10^{8.5}$\msun), the early- and late-type samples have essentially the same average index of $\sim0.7$. At larger masses, the known bi-modality of S\'{e}rsic index is seen between the late-type and early-type galaxies with the early-type galaxies being much more concentrated \citep[e.g.][]{bershady2000, goto2003}.

\begin{figure*}
\includegraphics[width=0.98\textwidth]{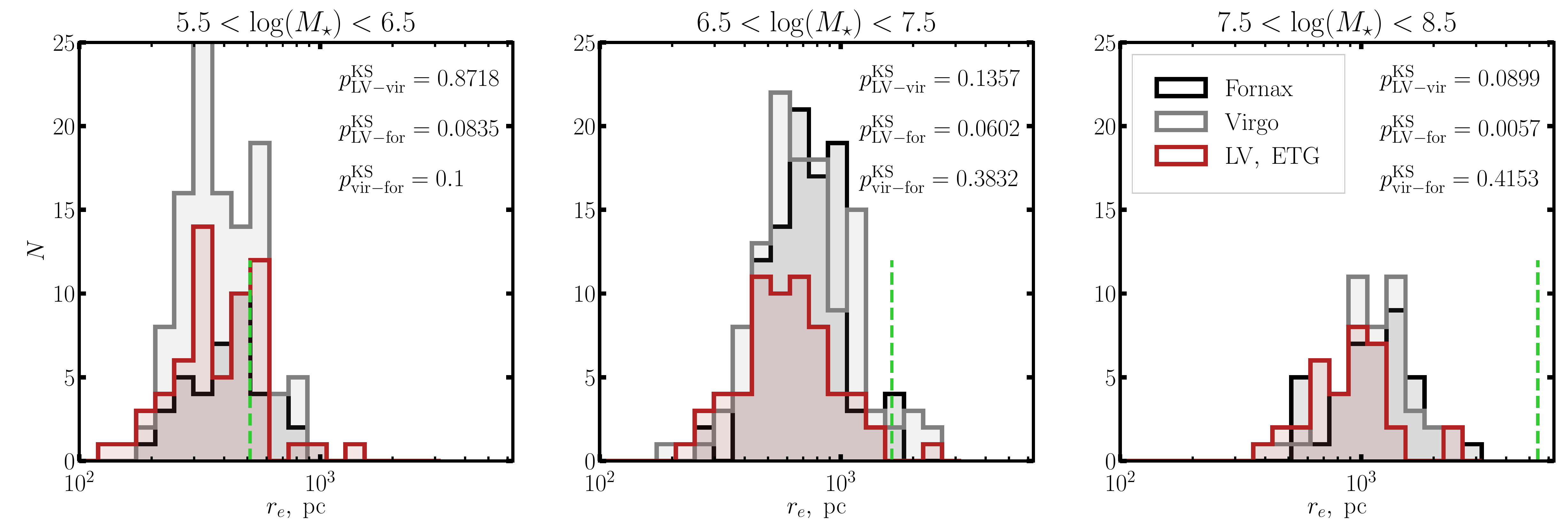}
\caption{The distribution of early-type dwarf sizes in three different bins of stellar mass. In each case the distribution of sizes at fixed stellar mass is roughly lognormal. The dashed green lines show the effective radii of dwarfs with stellar mass in the center of each bin and $\mu_{0,V}=26.5$ mag arcsec$^{-2}$, the surface brightness completeness limit of the ELVES Survey (assuming $n=0.7$ and $M_\star/L_V=1.2$). The completeness limit is amongst the observed dwarfs in the lowest mass bin, but we estimate the fraction of dwarfs missed due to low surface brightness is $\lesssim20$\%.}
\label{fig:size-dists}
\end{figure*}

\subsection{Size Distributions}
\label{sec:sizes}
In addition to the average mass-size relation, the distribution of sizes at fixed stellar mass is an interesting feature to look at for the different dwarf samples considered here. This has the dual purpose of 1) showing whether the distributions are the same between the cluster and LV samples and 2) of quantifying how many satellites with $M_V<-9$ might the ELVES Survey be missing with its surface brightness limit of $\mu_{0,V}<26.5$ mag arcsec$^{-2}$. From Figure \ref{fig:etg_v_ltg_lv}, it appears that the average scaling relation reaches $\mu_{0,V}\sim26.5$ mag arcsec$^{-2}$ right around $M_V\sim-9$, but since there is scatter, a number of dwarfs will have $\mu_{0,V}>26.5$ mag arcsec$^{-2}$ even with $M_V<-9$. The MW has no such dwarf satellites, but M31 has a few. The other nearby hosts that can be probed to very faint SB levels through resolved star searches of satellites (CenA and M81) also each have a few. 

Figure \ref{fig:size-dists} shows the distribution of sizes in three different 1 dex wide bins of stellar mass. For clarity, only the two cluster samples and the LV early-type satellite sample are shown. In each mass bin, the distribution of stellar size appears to be roughly lognormal, with the mean size increasing with stellar mass. Two-sample KS test $p$-values comparing the three samples in each mass bin are given in each panel. The larger sizes of the cluster dwarfs are visible most clearly in the high stellar mass bin.

Also shown in the panels of Figure \ref{fig:size-dists} are lines indicating the effective radius of a dwarf with a stellar mass in the center of each mass bin and central surface brightness of $\mu_{0,V}=26.5$ mag arcsec$^{-2}$, the completeness limit of ELVES. A mass-to-light ratio of $M_\star/L_V=1.2$ and S\'{e}rsic index of 0.7 are assumed in this calculation. Note that there are galaxies larger than this supposed limit in each of the panels because some of the real galaxies have different mass-to-light ratios or higher S\'{e}rsic indices. This limit is far above the locus of real dwarfs in the most massive bin but is within the scatter of observed galaxies in the lowest mass bin, indicating that ELVES is likely missing dwarfs due to the surface brightness limit.

To quantify the fraction of dwarfs lost, we integrate a Gaussian with mean given by the mean log effective radii and spread given by the standard deviation of the log effective radii from negative infinity up to the size limit imposed by the surface brightness limit. We find that we are missing $\sim20$\% of dwarfs in the lowest mass bin and $\sim1$\% in the intermediate mass bin. This will be an underestimate of the true fraction lost. The standard deviation of the distribution is shrunk by the fact that some larger galaxies are missed because they are too low surface brightness. If this was very severe, there would be a clear asymmetry with a sharp drop-off of dwarf abundance at large sizes, yet the distribution is largely symmetric, even in the lowest mass bin. Additionally, we get a similar loss fraction in the lowest mass bin if we use the observed scatter in the intermediate bin. With all this said, it does appear likely that ELVES is only missing $\lesssim20$\% of dwarfs in the mass range $10^{5.5}<M_\star<10^{6.5}$ \msun~due to surface brightness incompleteness. This is roughly consistent with the estimate from considering the subpopulation of LV satellites around the MW, M31, CenA, and M81 where satellites of much fainter surface brightness are discoverable via resolved stars. Additionally, using the extremely deep HSC data ($\sim$10 hours of integration on Subaru) from \citet{tanaka2017} for one ELVES host (NGC 4631), we do not find any additional satellites in the ELVES completeness range\footnote{Note, however, that we are able to detect a new satellite with $M_V\sim-7$ mag and $\mu_{0,V}\sim27.5$ mag arcsec$^{-2}$.}.

\subsection{Fitting a Mass-Size Relation}
\label{sec:msr}

In this section, we fit power laws to the mass-size relations of the various dwarf samples considered. For these fits, we focus on dwarfs in the mass range $5.5 < \log(M_\star/{\rm M}_\odot) < 8.5$. The lower limit is roughly the luminosity limit ($M_V=-9$ mag) of the survey and the upper bound is set so that we avoid many of the highest mass satellites which are invariably poorly fit by single S\'{e}rsic profiles.

To explore quantitatively how well the mass-size relations of the various dwarf samples agree with each other, we fit the mass-size relations of each group with power laws. Specifically, we fit a linear relation between $\log(r_e)$ and $\log(M_\star/M_\odot)$ of the form:
\beq 
\log(r_e/{\rm pc}) = a + b\log(M_\star/{\rm M}_\odot).\\
\eeq
Since there are significant errors on both $r_e$ and $M_\star$ for the dwarfs, we use the \texttt{linmix} algorithm to robustly fit a line \citep{kelly2007}\footnote{We use the \texttt{python} implementation of \texttt{linmix} written by J. Meyers: \url{https://github.com/jmeyers314/linmix}}. The \texttt{linmix}  algorithm uses Markov Chain Monte Carlo (MCMC) to sample the posterior distributions of the parameters of the linear regression, including intrinsic scatter. We assume that the errors in $\log(r_e)$ and $\log(M_\star/M_\odot)$ are Gaussian and uncorrelated. We take uniform priors on the slope, normalization, and intrinsic scatter squared. From the posterior distributions, we find median parameters as given in Table \ref{tab:fits} along with $1\sigma$ uncertainties. In this Table, $\sigma$ gives the intrinsic scatter in $\log(r_e)$ in dex. The uncertainties in this table are from marginalizing over the posterior distributions. Note that since the $a$ and $b$ are clearly covariant, these marginalized uncertainties will be somewhat overestimated. When plotting and analyzing these relations throughout the paper, we will incorporate uncertainties directly by sampling from the posterior distributions.

\begin{deluxetable}{cccc}
\tablecaption{Results of fitting the mass-size relation for different samples of dwarfs in the mass range $5.5 < \log(M_\star/{\rm M}_\odot) < 8.5$. $\sigma$ denotes the intrinsic scatter in logarithmic size at fixed stellar mass.\label{tab:fits}}
\tablehead{\colhead{Sample} & \colhead{$a$} & \colhead{$b$} & \colhead{$\sigma$}}
\startdata
LV, all  &  $1.071^{+0.127}_{-0.123}$  &  $0.247^{+0.018}_{-0.018}$  &  $0.181^{+0.01}_{-0.009}$  \\
LV, ETG  &  $1.179^{+0.155}_{-0.156}$  &  $0.231^{+0.023}_{-0.023}$  &  $0.183^{+0.012}_{-0.011}$  \\
LV, LTG  &  $0.802^{+0.249}_{-0.243}$  &  $0.283^{+0.034}_{-0.035}$  &  $0.192^{+0.019}_{-0.017}$  \\
Field  &  $0.667^{+0.199}_{-0.203}$  &  $0.296^{+0.028}_{-0.028}$  &  $0.215^{+0.015}_{-0.013}$  \\
Cluster  &  $1.027^{+0.082}_{-0.084}$  &  $0.259^{+0.012}_{-0.012}$  &  $0.152^{+0.006}_{-0.006}$  \\
\enddata
\end{deluxetable}

\begin{figure*}
\includegraphics[width=0.98\textwidth]{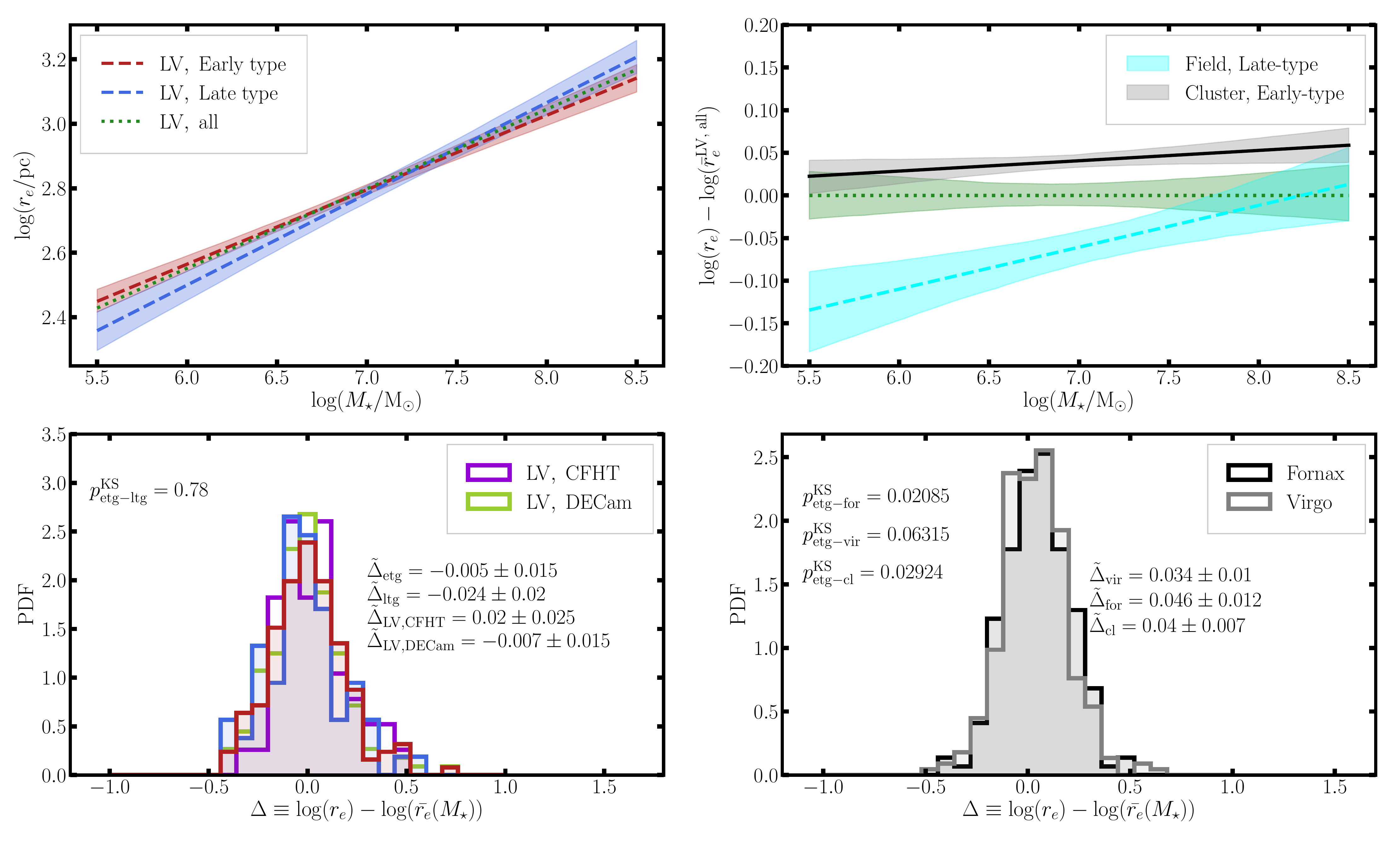}
\caption{\textit{Top}: The left panel shows the mass-size relations for three groupings of LV satellites (early-type-red, late-type-blue, all-green). The right panel shows the relations of the cluster (grey) and field (cyan) dwarf samples normalized to the full LV sample. At fixed stellar mass, the cluster sample is offset to larger sizes whereas the isolated field sample is offset to smaller sizes. For the latter, however, the sample might be incomplete to larger, lower surface brightness dwarfs, causing the offset. \textit{Bottom}: Distributions of the residuals in $\log(r_e)$ from the best-fit mass-size relation of the full LV satellite sample are shown for the various samples. The LV satellites are also split by photometry source (CFHT/Megacam and DECaLS/DECam). $\tilde{\Delta}$ indicates the median residual for the various groups. $p$-values from two-sample KS tests comparing the residual distributions from various subsamples are also given in each panel.}
\label{fig:mass-size-resids}
\end{figure*}

These relations are shown in Figure \ref{fig:mass-size-resids}. As qualitatively indicated by Figure \ref{fig:etg_v_ltg_lv}, we find that the mass-size relations from the late-type and early-type LV satellite samples are quite similar. The late-type relation has a slightly steeper slope but it is consistent with that of the early-type satellites within the errors (upper left panel). In the top right panel, we show the various mass-size relations normalized by the best-fit relation for the complete LV satellite sample. The field sample is smaller at fixed stellar mass, particularly at low masses. As we have said before, due to the unclear surface brightness completeness of this sample, this might be due to observational incompleteness and not a physical discrepancy. On the other hand, the cluster sample, which does have well-quantified completeness, is larger at fixed stellar mass than the LV satellites at the $\sim2-3\sigma$ level. 

In the bottom row of Figure \ref{fig:mass-size-resids}, we show the residuals of the mass-size relations using the best-fit relation for the full LV sample.  We see that the sizes of the LV ETGs and LTGs are consistent within $\sim0.025$ dex ($\sim5$\%) once corrected for the overall mass-size relation. A two-sample KS test indicates that the two distributions are not significantly different ($p$-value given in the figure panel). Recalling that the CFHT data is generally substantially deeper than the DECam/DECaLS data, the LV satellite sample is also split into `CFHT' and `DECam' satellites based on the source of their photometry, in order to check for possible systematic biases between these sources.  The `CFHT' and `DECam' LV satellites are indistinguishable in size, given the uncertainties. 

The bottom right panel shows the residuals for the two cluster satellite samples, again using the best-fit relation for the full LV sample. Given the greater than an order of magnitude difference in host halo mass of these dwarf samples (cluster vs. LV), their similarity in average size at fixed stellar mass is striking. With that said, however, the cluster residuals are biased from zero in the sense that the cluster satellites are $\sim0.04$ dex (8\%) larger than the LV satellites when controlling for stellar mass. The two-sample KS tests indicate that the residuals are significantly different from the distribution of residuals for the LV early-type satellites, particularly the Fornax and combined sample. Our fiducial analysis is for dwarfs in the mass range $5.5 < \log(M_\star/{\rm M}_\odot) < 8.5$, but we find a similar offset in size considering slightly different mass limits. 

We remind the reader that the cluster satellite sizes and stellar masses are measured in \textit{exactly the same way} as the LV satellites. Furthermore, the LV satellite sample is roughly split between measurements taken with CFHT/Megacam (using $g/i$ and $g/r$) and Blanco/DECam (using $g/r$) while the Virgo cluster sample uses CFHT/Megacam (using $g/i$) and the Fornax sample uses Blanco/DECam (using $g/r$). Thus it is difficult to imagine a way in which telescope/filter differences could conspire to create the offset in sizes.

\section{Dwarf Intrinsic Shapes}
\label{sec:shapes}
In addition to the projected size and surface brightness of dwarfs, their intrinsic three-dimensional shapes hold important clues to their formation and the physical processes relevant in their evolution. In particular, in this section we use the distribution of observed (projected) ellipticities to infer the average intrinsic axis ratios of the dwarf populations. The intrinsic axis ratios are sensitive to the effects of stellar feedback \citep[e.g.][]{kaufman2007, rsj2010}, gas content and morphology \citep[e.g.][]{roychowdhury2010}, and tidal stripping \citep[e.g.][]{lokas2012, barber2015}. Stronger feedback and/or more tidal stripping will generally lead to rounder galaxies. Since the sizes are also sensitive to these processes, this is an important alternative perspective on the results we found in the last section on the mass-size relation amongst the different dwarf samples. 

The two LV satellite samples (early-type and late-type) show indication that they have different intrinsic shapes just from their observed ellipticity distributions. The late-type dwarfs have average ellipticity (within the stellar mass range of $10^{5.5}<M_\star<10^{8.5}$\msun) of 0.40 while the early-types have an average of 0.29. A two-sample KS test indicates the distributions are significantly different with a $p$-value of $5\times10^{-4}$. From Figure \ref{fig:etg_v_ltg_lv}, it is clear that the LTG sample is biased to higher luminosities and masses. To disentangle this effect from the effect of galaxy type on ellipticity and intrinsic shape, we derive a mass-matched ETG subsample for more direct comparison with the LTG sample. Essentially, we pare down the larger ETG sample by randomly removing galaxies using the LTG sample's mass distribution to determine the probability of keeping each ETG. This `mass-matched' ETG sample has a higher average stellar mass than the overall ETG sample. Interestingly, the average ellipticity of this subsample is 0.30, similar to the overall ETG sample. The ellipticity distribution of this subsample is also significantly different from that of the LTG sample with a $p$-value of 0.004 from a KS test. For most of the following results, we use the entire ETG sample but also consider the mass-matched ETG subsample as an additional check. 

Due to the fact that the orientation of a given galaxy with respect to the observer is unknown without kinematic information, the intrinsic axis ratios can only be inferred for a population, assuming a random distribution of viewing orientations. We assume that the dwarfs are represented by a family of possibly triaxial ellipsoids, described by the three axis lengths: $C<B<A$. The intrinsic shapes are then given by the ratios of short to long axis and intermediate to long axis, $C/A$ and $B/A$. This approach has been taken numerous times in the literature for a wide variety of galaxy samples \citep[e.g.][]{lisker2007, padilla2008, roychowdhury2013, rsj2010, rsj2016, rsj2019_shapes, salomon2015, sanders2017, kadofong2020}. Most relevant to the current work, these earlier works find that early-type dwarfs (both in clusters and in the Local Group) are well represented by roughly oblate spheroids with intrinsic axis ratios $C/A\sim0.5-0.6$. In particular \citet{rsj2016,rsj2019_shapes} inferred from the NGVS dwarf sample an intrinsic axis ratio of $C/A=0.57$. They found that the Local Group (LG) dwarfs had a similar axis ratio of $C/A=0.49$. 

To infer the intrinsic axis ratios from the observed distribution of ellipticities, we largely follow the procedure of \citet{rsj2016} and \citet{kadofong2020}. We assume the underlying population of dwarfs is a family of optically thin ellipsoids, described by a mean intrinsic ellipticity, $\bar{E}$, and triaxiality, $\bar{T}$, with intrinsic (Gaussian) dispersions, $\sigma_E$ and $\sigma_T$. In terms of intrinsic axis ratios, the intrinsic ellipticity and triaxiality are given by: $E=1-C/A$ and $T=(A^2-B^2)/(A^2-C^2)$. We use a Bayesian framework to infer $\bar{E}$, $\bar{T}$, $\sigma_E$, and $\sigma_T$ from the distribution of observed ellipticities, $\epsilon=1-b/a$.

We infer the intrinsic ellipticity and triaxiality via MCMC. For each iteration, we predict a distribution of observed ellipticities by considering $10^5$ random draws of $E$ and $T$ combined with a random sight-line. $E$ and $T$ are both limited to the range [0,1] so we use a truncated normal distribution (with means of $\bar{E}$/$\bar{T}$ and dispersions $\sigma_E$/$\sigma_T$) to draw samples. We express the observed ellipticity in terms of the intrinsic axial ratios and the viewing angles using the equations of \citet{binney1985}. This distribution of predicted ellipticities is convolved with a Gaussian with dispersion 0.08 to represent a typical observational uncertainty for ellipticity in the ELVES sample (see Appendix \ref{app:uncertainty}). The distribution of predicted ellipticities is then binned and compared to the binned distribution of observed ellipticities from the ELVES Survey using the standard Poisson likelihood function:
\beq
\ln \mathcal{L}(\epsilon | \bar{E},\bar{T}, \sigma_E, \sigma_T) = \sum_i n_i \ln(m_i) - m_i - \ln(n_i!) \\
\eeq
where $n_i$ is the observed count in bin $i$ and $m_i$ is the predicted count. Assuming flat priors in $\bar{E}$ and $\bar{T}$, we then use MCMC to sample from the posterior of $\bar{E}$, $\bar{T}$, $\sigma_E$, and $\sigma_T$. We use bin sizes of 0.05, but the results are not dependent on this. If a given set of parameters leads to a bin having a predicted count of 0, a filler value of 0.001 is used instead. 

We have checked our method using simulated samples of observed ellipticities. In these tests, we generate a mock observational sample with sample size of 200 using a specific set of $\bar{E}$ and $\bar{T}$ (and setting $\sigma_E=\sigma_T=0.1$).  For each mock dwarf, $E$ and $T$ are drawn from a truncated normal distribution and the observed ellipticity along a random sight-line is calculated, assuming an observational error of 0.08 in the ellipticity. We then use the MCMC-based inference method described above to infer the underlying $\bar{E}$ and $\bar{T}$. We find that our method can readily recover the intrinsic axis ratios within the uncertainties for a wide range of $\bar{E}$ and $\bar{T}$.

\begin{deluxetable}{ccccc}
\tablecaption{Results of the intrinsic shape analysis.\label{tab:shapes}}
\tablehead{\colhead{Sample} & \colhead{$n_{\rm gals}$} & \colhead{$\langle M_V\rangle$} & \colhead{$\langle B/A \rangle$} & \colhead{$\langle C/A \rangle$}}
\startdata
LTG, all & 66 & $-13.51\pm1.65$ & $0.78\pm0.15$  & $0.35\pm0.10$ \\
ETG, all & 154 & $-11.76\pm1.62$ & $0.85\pm0.12$ &  $0.55\pm0.03$ \\
\enddata
\end{deluxetable}

\begin{figure*}
\includegraphics[width=\textwidth]{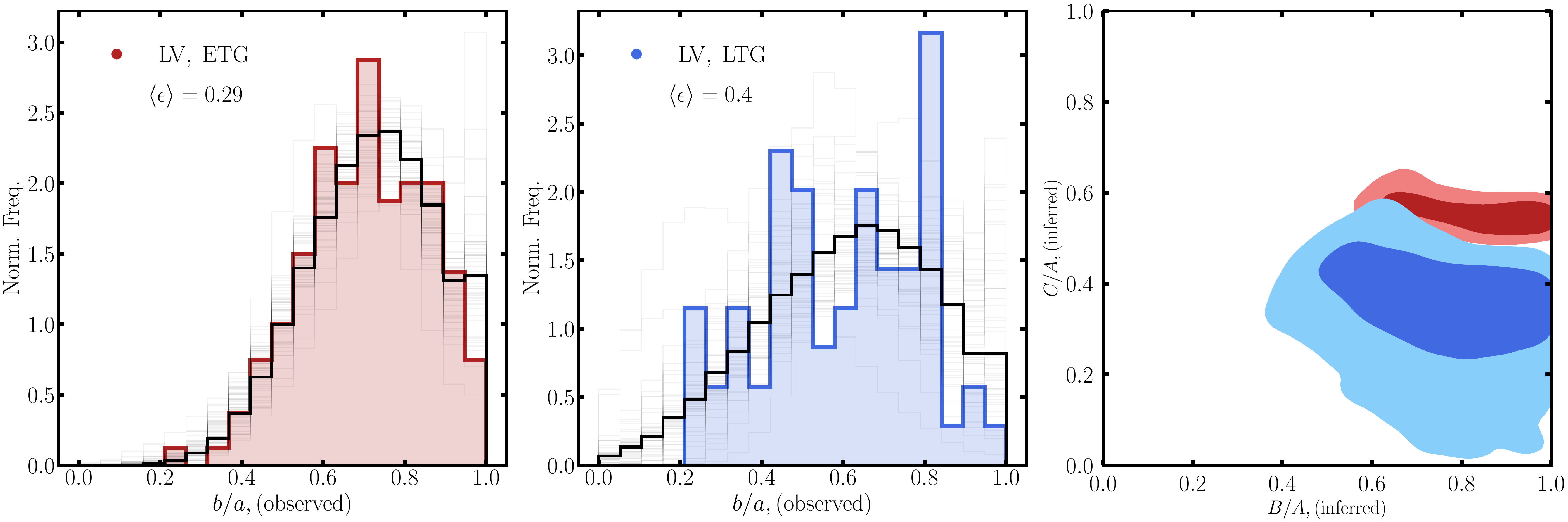}
\caption{\textit{Left and Center}: The observed distributions in apparent axial ratio ($b/a=1-\epsilon$) for the early-type and late-type LV satellites are shown in the color histograms. Black histograms show the median model from the Bayesian inference of the intrinsic axial ratios while the thin gray lines show individual draws from the posterior distributions.  \textit{Right}: $1$ and $2\sigma$ contours showing the inferred, intrinsic $C/A$ and $B/A$ regions for the LV early and late-type samples. }
\label{fig:shapes_inf}
\end{figure*}

\begin{figure*}
\includegraphics[width=\textwidth]{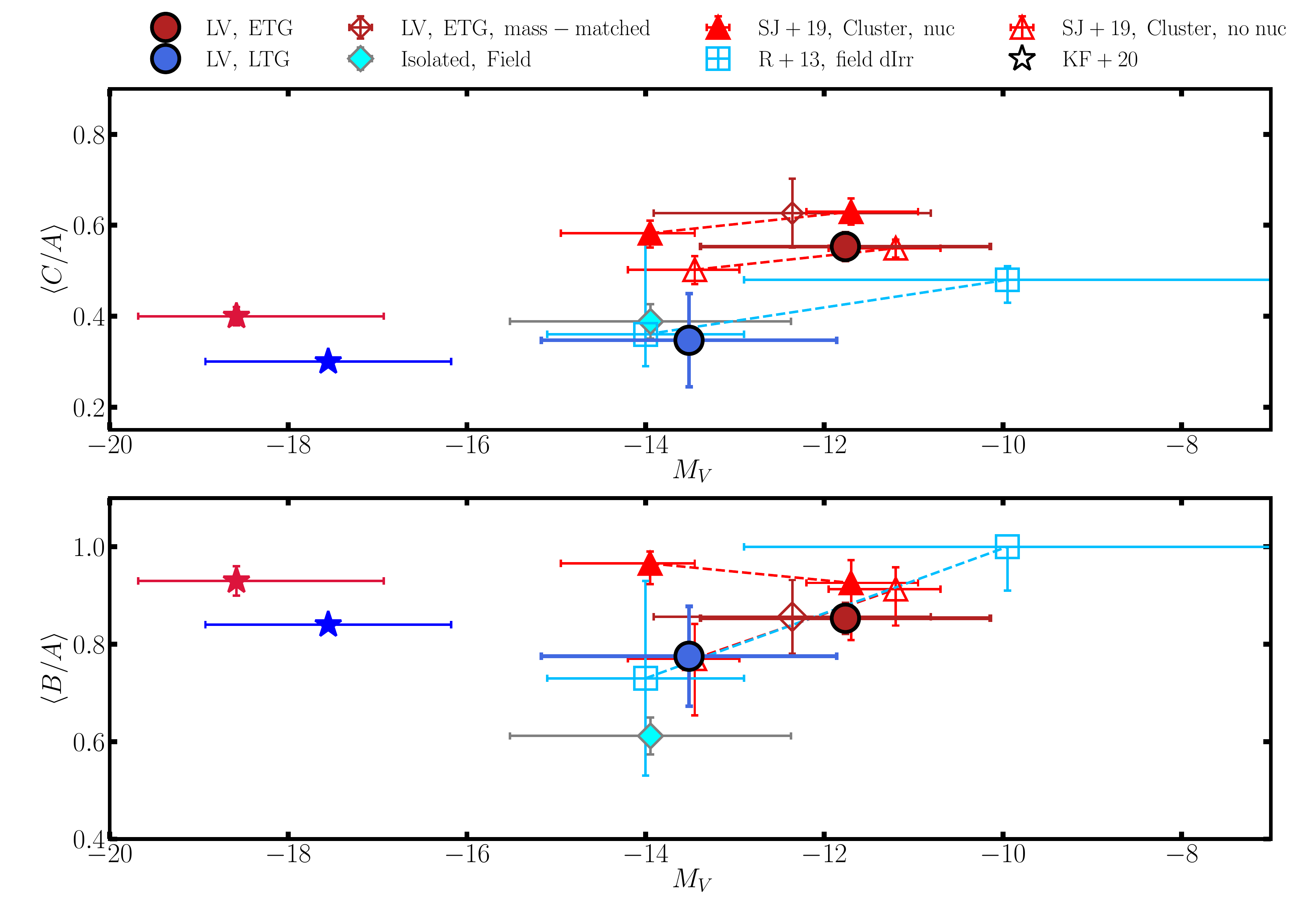}
\caption{Inferred average intrinsic axial ratios: $C/A$ (top) and $B/A$ (bottom), for different dwarf samples as a function of the average luminosity of the samples. The `mass-matched' ETG sub-sample has been selected to have the same distribution of stellar masses as the late-type satellite sample. The cluster sample analyses come from \citet{rsj2019_shapes}, the field dIrr analysis comes from \citet{roychowdhury2013}, and we also include the analysis of higher mass dwarfs from \citet{kadofong2020}. Dashed lines connect the points from studies with two luminosity bins.}
\label{fig:shapes}
\end{figure*}

The results are given in Table \ref{tab:shapes} for the ELVES sample for stellar masses in the range $10^{5.5}<M_\star<10^{8.5}$\msun~ and shown in Figure \ref{fig:shapes_inf}. The dwarfs are generally oblate with $B/A\sim1$, although the late-type dwarfs appear to be mildly triaxial. The average thickness of the early-type dwarfs of $\sim0.55$ agrees quite closely with the average thickness of the NGVS sample of $0.57\pm0.02$ found by \citet{rsj2016}. The late-type dwarfs are noticeably more flattened with $C/A\sim0.3$ \footnote{Note that Table \ref{tab:shapes} gives the results for the complete sample of LV ETGs (i.e. not the mass-matched subsample). The intrinsic shape results for both the full and mass-matched ETG samples are shown in Figure \ref{fig:shapes}.}.

In a combined analysis of Virgo and Fornax dwarfs, \citet{rsj2019_shapes} found that intrinsic thickness was a function of both dwarf luminosity and nucleation with fainter and nucleated dwarfs being intrinsically thicker. Due to the relatively small sample size of the ELVES Survey with respect to the combined cluster sample, we do not attempt to split the satellite sample into luminosity or nucleation bins. Still, we compare how the LV satellite samples compare with the different sub-samples considered by \citet{rsj2019_shapes} in Figure \ref{fig:shapes}, comparing both $C/A$ and $B/A$. Also plotted are the results from \citet{kadofong2020} for higher mass dwarfs\footnote{Note that \citet{kadofong2020} splits between early- and late-type based on a color cut of $g-i=0.9$ while the other samples in this plot are split primarily on morphology.} and the results from anaylzing dIrrs from the Nearby Galaxy Catalog \citep{karachentsev2013} of \citet{roychowdhury2013}. Additionally, we include an analysis of our `field isolated' late-type sample. This sample will naturally have large overlap with the \citet{roychowdhury2013} sample since both are based on the catalog of \citet{karachentsev2013}, although we use more modern photometry and a slightly different method. It is reassuring that they agree quite well. It appears that at least some of the difference in $C/A$ between ELVES early-type and late-type dwarfs is attributable to the difference in average luminosity and stellar mass. However, the late-type sample is significantly flatter than the mass-matched early-type sample, so this cannot be all or even most of it.

The top panel of Figure \ref{fig:shapes} shows that the early-type and late-type dwarfs exhibit a roughly constant offset in intrinsic $C/A$ across the probed range in luminosity. Both types of galaxies appear to get thicker at lower luminosities, but they show a difference in intrinsic shape at all luminosities.

The bottom panel of Figure \ref{fig:shapes} shows that there is no similar clear pattern in intrinsic $B/A$. Other than the faintest point of \citet{roychowdhury2013}, it appears that early types have higher $B/A$ (less triaxial) than late types at all masses. Also the three late-type data points around $M_V\sim-14$ mag have lower $B/A$ (more triaxial) than the higher-mass, late-type data point of \citet{kadofong2020}, indicating a possible mass trend \citep[again ignoring the faintest bin of ][]{roychowdhury2013}. Other than these points, it is difficult to draw conclusions from the inferred $B/A$'s.

In the preceding section, we found that overall the early and late-type ELVES satellites had quite similar structures, once the different stellar populations were accounted for. Here, however, we find that there is a definitive difference in structure between the two populations. The late-types are significantly more flattened. Additionally in the last section, we found that the cluster satellites were somewhat larger at fixed stellar mass than the LV early-type satellites. However, here we do not find a significant difference in intrinsic shapes between cluster satellites and the LV early-types. We explore what these similarities and differences in structure between the different dwarf samples might mean for dwarf galaxy evolution in the next section.

\section{Discussion}
\label{sec:disc}
Thus far in this paper, we have explored various aspects of the structure of low-mass, dwarf galaxies ($M_\star\lesssim10^9$\msun), comparing dwarfs of different types (late- vs. early-type) and in different environments (cluster environments vs in the halos of MW-like galaxies). Overall, there is striking similarity in the scaling relations of dwarfs across types and environments, indicative of both an evolutionary tie between late- and early-type galaxies and that environment plays a relatively weak role in determining the structure of dwarf galaxies. In this section, we delve deeper into each of these points, examining what each could mean for dwarf galaxy evolution.

\subsection{Late-type vs. Early-type Dwarf Structure}
\label{sec:ltg_vs_etg}
There is significant current consensus that dwarf early-type galaxies were once late-type dwarfs of similar mass that have been quenched via ram pressure stripping  \citep[e.g.][]{boselli2014, boselli2014_review, fillingham2015, fillingham2016, venhola2019, janz2021}. Ram pressure stripping is invoked due to the need for a very rapid quenching mechanism (operating on timescales $\lesssim1$ Gyr) to explain the significant absence of star-forming dwarfs in the centers of clusters and also their paucity in the halo of the MW and M31. Of course, the transformation of a late-type dwarf into an early-type involves more than just quenching; it also involves any necessary morphological change. However, there is additional growing consensus that a drastic morphological change is simply not needed. In terms of kinematics, dwarf early-type galaxies overall show fairly similar levels of rotation support compared to star forming dwarfs of similar mass \citep{toloba2009, toloba2011, wheeler2017}. These lines of evidence suggest that the transformation in many cases is just tantamount to the removal of gas and cessation of star-formation.

The similarity in sizes (Figure \ref{fig:mass-size-resids}) and S\'{e}rsic indices (Figure \ref{fig:sersic_n_all}) between late-type and early-type LV satellites seems to fit right into this picture. However, our results go further and show that a drastic morphological change really \textit{cannot} occur. The late-type and early-type satellites have the same sizes at fixed mass within 5\%, and the S\'{e}rsic indices are similarly indistinguishable for $M_\star<10^{8.5}$\msun. This is an important constraint on numerical models of this transformation process. Models of this transformation process that involve significant mass-loss either due to gravitational harassment in a cluster \citep[e.g.][]{mastropietro2005} or tidal stirring by a MW-like host \citep[e.g.][]{mayer2006, kazantzidis2011} would seem unlikely to maintain such a tight similarity between the structure of the early-type dwarf and its late-type progenitor. In general, simulations of tidal stripping and heating show that dwarf spheroidals grow as they are stripped \citep[e.g.][]{penarrubia2008, errani2015, errani2018}, although this depends on the shape of the underlying dark matter halo. In essence, in this work we have quantified the structure of the starting point and ending point for this transformation process, and it will be important to see if simulations of dwarf quenching via ram pressure stripping, including any tidal evolution, can connect the two. 

Observations indicate that there is likely some continual tidal evolution of early-type satellites after they have been quenched. \citet{toloba2015} found that dwarf ellipticals in Virgo were more likely to be slow rotators in the center of the cluster as opposed to at the outskirts \citep[see also][for results in Fornax]{scott2020}. Since galaxies on the outskirts have likely experienced more recent infall than galaxies in the center (on average), this would be indicative of an evolutionary effect, with dwarfs kinematically heating up as they orbit in the cluster. However, the similarity in sizes between late-type and early-type dwarfs and also the fact that early-type dwarfs in Fornax do not show a strong dependence of size on position in the cluster \citep{venhola2019} would indicate that this kinematic evolution does not involve significant size evolution.

An additional complication in comparing size evolution in different environments is the fact that the late-type dwarfs are likely evolving in size as they continue forming stars in isolation. Studies of high-redshift galaxies \citep[e.g.][]{vanderwel2014} show that the mass-size relation of galaxies evolves with redshift, although the evolution is less steep for star-forming galaxies and also for lower mass galaxies \citep{mowla2019}. These studies are focused on galaxies with $M_\star > 10^9$ \msun, so it is unclear what is happening on the dwarf scales considered in this work. A ubiquitous prediction of modern hydrodynamic simulations with `bursty' star formation is that stellar orbits can be heated, causing the galaxy to grow \citep{elbadry2016, dicintio2017, chan2018}\footnote{These models predict galaxies to grow as they form stars (in other words, both stellar mass and radius increase), thus it is unclear what they predict the evolution of the mass-size relation to be.}, analogous to DM core formation. Thus, it is possible that the progenitors of today's early-type dwarf satellites were smaller than today's late-type dwarfs, requiring significant size growth in the transformation process to balance out the growth of the late-type dwarfs due to star formation feedback. A perfect balancing between the two effects seems unlikely and coincidental if both were drastic, and we assert it is more likely that the size evolution involved in both the creation of an early-type satellite and in star formation feedback is quite mild. This tight correspondence between late-type and early-type dwarfs could be a useful constraint on these models of feedback.

However, in the above picture, it is difficult to reconcile the results on the intrinsic shapes of late- versus early-type dwarfs. Figure \ref{fig:shapes} \citep[see also][]{rsj2019_shapes} shows that early-type and late-type dwarfs show a significant difference in shape at all dwarf masses probed. This indicates that at least \textit{some} structural evolution is still required in the creation of early-type satellites, at odds with the tight similarity found between other measures of dwarf structure. In the above paragraph we noted that it was likely that late-type dwarfs are growing mildly with time as they form stars in isolation which allows some room for growth due to tidal stripping/heating for early-type dwarfs after being quenched as they fall into their host halos. Perhaps this mild evolution in size is accompanied by a significant thickening of the stellar distribution (from intrinsic $C/A\sim0.35$ to $\sim0.55$). To understand if this is possible, simulations that track both the evolution of size and intrinsic thickness over the entire range of dwarf masses probed here are needed. To our knowledge, this has not been well-explored beyond the fact that stripped dwarfs get thicker \citep[e.g.][]{kazantzidis2011,lokas2012,barber2015,tomozeiu2016}. Perhaps it is possible to derive tidal tracks for intrinsic axial ratios like \citet{errani2018} do for satellite size and mass which would allow for more detailed comparison of the evolution of dwarf size with intrinsic shape as stripping occurs.

Alternatively, it is possible that the thicker intrinsic shapes of early-type dwarfs comes from passive fading of young stars recently formed in the disks of their gas-rich, late-type progenitors. Dwarf galaxies are known to ubiquitously host round stellar halos \citep[e.g.][]{kadofong2020}, and as the young, bright stars in a recently-quenched dwarf galaxy (which will naturally form in a flattened configuration) passively fade, the rounder, older outskirts of the galaxy will become more apparent. Models that track the size and intrinsic shape evolution of dwarfs during the late-type to early-type transition will need to account for this. With the current data, it is unclear which effect (passive fading or tidal stripping/heating) would be more dominant in changing the intrinsic shapes of quenching dwarfs. We note that in the fading scenario, the sizes might evolve as well, running into a similar problem as the tidal stripping/heating scenario described above. Additionally, the similarity in S\'{e}rsic index (both have $n\sim1$) could be another constraint on this passive fading scenario.

\subsection{Cluster vs. Local Volume Satellites}
The second main comparison we did in this paper is between the early-type satellites in the Virgo and Fornax clusters and early-type satellites of much lower mass Local Volume hosts. Overall we found that the LV satellites exhibited very similar scaling relations as the cluster satellite sample, at least to about the $\sim10$\% level. At first brush, given the significant difference in host halo masses, this indicates that environment plays a pretty minor role in dwarf structure. This is seen in dwarf sizes, surface brightness, and intrinsic shapes.

With the sample size of LV satellites afforded by the ELVES Survey, we are able to identify some subtle, but robust, differences between the cluster and LV satellite samples. Mainly, we find that the cluster satellites are larger than the LV satellites at fixed stellar mass by $\sim10$\%. We interpret this as evidence of increased tidal heating of cluster satellites compared to the LV satellites. While most cluster satellites are not stripped to the point that they start to lose stellar mass \citep[e.g.][]{smith2015}, many will have some significant amount of their DM mass stripped. Simulations of stripped dSphs indicate that usually $\gtrsim90$\% of the satellite's DM halo must be stripped before it starts to lose stars \citep[e.g.,][]{penarrubia2008, penarrubia2010}. 

Due to their generally earlier infall\footnote{Note that the current NGFS and NGVS cluster samples are in the \textit{cores} of their respective clusters, accentuating this difference.} and more extreme environment, we expect the cluster satellites to have higher fractions of their DM mass stripped than the LV satellites. Considering the tidal tracks of \citet{errani2018}, which predict the evolution of the stellar mass and half-light radius of satellites as a function of the fraction of their DM that has been stripped, we would expect the cluster satellites to be somewhat larger, as is observed (Figure \ref{fig:mass-size-resids}). However, this is complicated again by the fact that the size distribution of late-type progenitors for the cluster satellites could be different than that of the LV satellites since the late-type mass-size relation likely evolves with redshift, and the cluster satellites will have experienced infall earlier. Additionally, there is uncertainty whether the cluster satellites obey the same stellar to halo mass relation as the LV satellites \citep[e.g.][]{grossauer2015}. Addressing these complications to get a robust quantitative prediction for the size of cluster satellites compared to LV satellites is out of the scope of this work, but we will pursue it in the future.

\subsection{Structural Changes at a Characteristic Mass Scale of $M_\star\sim10^{8.5}$\msun}
\label{sec:ram pressure stripping}

In this paper, we have found that the tight similarity in size and S\'{e}rsic index between late-type and early-type dwarfs only holds for dwarfs with $M_\star\lesssim 10^{8.5}$ \msun ~(Figure \ref{fig:sersic_n_all}). For higher mass dwarfs, the early-type dwarfs are smaller and exhibit more concentrated light profiles. This transition, occurring at roughly the same transition mass of $M_\star\sim 10^{8.5}$ \msun~for both the mass-size relation and S\'{e}rsic indices, is suggestive that this is a physically meaningful scale. The fact that ETGs above this mass scale are smaller and more centrally concentrated than similar mass LTGs currently existing in the field indicates that these ETGs are not simply ram-pressure stripped (and, hence, quenched) field LTGs that have fallen into the cluster. Note that there are essentially no early-type LV satellite dwarfs above this $M_\star\sim 10^{8.5}$ \msun~mass scale, indicating that quenching and morphological transformation is altogether inefficient in MW-like halos above this mass \citep[see also][]{fillingham2015,fillingham2016,mao2020}. Their rarity in the field also indicates that their formation is greatly enhanced in denser environments. In other words, additional physical processes (likely endemic to the cluster environment) are required for the creation of these higher-mass dwarf ETGs.

We speculate that at least one process involved in the creation of these more compact ETGs is possibly dwarf-dwarf mergers in the gas-rich progenitors to these ETGs. The presence of compact ETGs above this mass scale could then be related to the emergence of kinematically cold disks in gas-rich dwarfs above roughly this mass. Both observations \citep[e.g.][]{rsj2010} and theory \citep[e.g.][]{dekel2020} indicate that thin disks are only present for dwarfs with stellar mass $M_\star\gtrsim 10^9$\msun. Dwarfs with stellar mass below this get systematically thicker (c.f. Figure \ref{fig:shapes}) and have hotter kinematics \citep[e.g.][]{wheeler2017}.  Mergers between disky dwarf galaxies can lead to very centrally concentrated remnants whereas mergers between spheroidal dwarfs are less likely to do so \citep[e.g.][]{bekki2008}\footnote{This is also related to the classic results that violent relaxation after mergers of disk galaxies can lead to de Vaucouleurs profiles \citep[e.g.][]{barnes1988, barnes1992}.}, thus explaining why early-type dwarfs above this mass scale can be centrally concentrated while lower-mass dwarfs are not. This is compounded by the fact that, in the cluster environment, the remnants will lose their gas quickly, with no time to reform a disk. 

This is largely conjecture, but it seems inescapable that more than just ram pressure stripping of field LTGs (like those that exist today) is required to create dwarf ETGs of masses $M_\star\gtrsim 10^{8.5}$\msun. It is likely that the gas-rich progenitors to these ETGs are unlike the LTGs we observe in the field currently. The high globular cluster specific frequency of many of these cluster early-type dwarfs corroborates this \citep{rsj2012}. It is possible the difference in progenitors is due to high merger rates in the cluster environment as it was forming. To investigate this further, it would be interesting to consider the kinematics of cluster dwarf ellipticals in the light of their structure. For instance, are the slow rotator dE's smaller and/or more centrally concentrated than the fast rotators?

\section{Conclusion}
\label{sec:concl}
In this paper, we explored the scaling relations between various structural parameters of dwarf satellites in the Local Volume, focusing primarily on the relation between stellar mass and galaxy size for dwarfs in the mass range $10^{5.5}<M_\star<10^{8.5}$ \msun. We use an unprecedented sample of low-mass and low-luminosity satellites from the ongoing ELVES (Exploration of Local VolumE Satellites) Survey that is surveying the dwarf satellites of \textit{all} massive hosts in the LV, down to luminosities of $M_V<-9$ and surface brightness $\mu_{0,V}<26.5$ mag arcsec$^{-2}$. We separate the satellite sample into late-type and early-type dwarfs based on visual inspection and argue that, based on available spectra and H\textsc{I}, this essentially corresponds to a split into star-forming and quenched dwarfs. Due to the fact that low-mass satellites ($M_\star \lesssim 10^{8.5}$ \msun) are expected to quench rapidly upon entering their host's halo \citep{fillingham2015, wetzel2015}, we argue that the late-type satellites are those satellites that are only just now falling into their host and, thus, have experienced mild, if any, processing by their host. We compare these two sub-samples with contemporary dwarf samples from the cores of the Virgo and Fornax clusters from the NGVS and NGFS Surveys. By comparing the dwarf structure between samples with different star formation histories and in different environments, we are able to disentangle the myriad of physical processes involved in sculpting dwarfs and extract insights into the relevant physics in dwarf galaxy evolution.

In this section, we provide an overview of the main takeaways of this paper. 

\begin{enumerate}
    \item The fraction of late-type dwarfs is a steep function of dwarf luminosity, reaching essentially zero at $M_V>-10$ mag (Figure \ref{fig:etg_v_ltg_lv}). We interpret this is due to the rapid quenching of low-mass satellites \citep{fillingham2015, wetzel2015, akins2020} due to ram-pressure stripping \citep{fillingham2016} from the host's hot gas halos. Lower mass satellites are less able to retain their gas in the face of ram pressure and are thus quenched even faster.
    
    \item Late-type dwarfs are, on average, slightly smaller and higher surface brightness than the early-type dwarfs at fixed luminosity. This appears to be an effect of different stellar populations. The difference in luminosity at fixed size can be explained by the passive ageing of a metal poor population (Figure \ref{fig:etg_v_ltg_lv}).
    
    \item We fit single power laws between galaxy size and stellar mass, including fitting for intrinsic scatter in this relation. The results are given in Table \ref{tab:fits}. The mass-size relations are very similar between the early-type and late-type LV satellite samples to within our observational uncertainty of $\sim5$\% in the mass range $10^{5.5}<M_\star<10^{8.5}$ \msun, although the late-type dwarfs are larger above this mass. The similarity in size indicates that the quenching and transformation of a late-type dwarf into a early-type dwarf involves only very mild size evolution.

    \item The mass-size relation of the cluster dwarfs is also strikingly similar to that of the LV satellites given the greater than an order of magnitude difference in host halo masses. With that said, the cluster dwarfs are larger at fixed stellar mass. We find a significant median difference of $\sim0.040\pm0.007$ dex ($\sim8$\%) at fixed stellar mass (Figure \ref{fig:mass-size-resids}). We argue this is due to increased tidal stripping and heating of satellites in the extreme cluster environments. 
    
    \item The reference sample of isolated field dwarfs from \citet{karachentsev2013} exhibit a mass-size relation that is quite similar to the LV late-type satellites. This similarity likely indicates that the late-type satellites have indeed not experienced any significant environmental processing. However, it is difficult to interpret as the field sample might be incomplete to very low-surface brightness dwarfs.
    
    \item The LV early-type satellites and cluster satellites both show similar, roughly lognormal distributions of sizes at fixed stellar mass (Figure \ref{fig:size-dists}). By considering the expected sizes of dwarfs with surface brightness below our fiducial completeness limit of $\mu_{0,V}<26.5$ mag arcsec$^{-2}$, we argue that ELVES is likely missing $\sim20$\% of dwarfs with stellar masses $10^{5.5} < M_\star < 10^{6.5}$ \msun~ due to low surface brightness but no dwarfs of higher stellar mass.
    
    \item The light profiles of the late-type and early-type satellites had very similar levels of concentration (measured by the S\'{e}rsic index) in the mass range $10^{5.5}<M_\star<10^{8.5}$ \msun, also indicating mild structural evolution in the transformation of a late-type dwarf into an early-type. Above this mass, the early-type dwarfs are significantly more centrally concentrated.

    \item Considering the observed distribution of apparent ellipticities, we infer the intrinsic shapes of the early-type and late-type LV satellite samples, assuming they are drawn from an underlying population of triaxial ellipsoids described by a mean intrinsic ellipticity, $\bar{E}$, and triaxiality, $\bar{T}$. Combining the intrinsic shapes of the ELVES dwarfs with samples from the literature, we find that both late-type and early-type dwarfs get thicker at fainter luminosities but early-type dwarfs are always distinctly rounder at fixed luminosity (Figure \ref{fig:shapes}).
    
\end{enumerate}

Overall these results motivate further comparison with simulations. In particular we note three key areas where further work with simulations would be illuminating. First, modern hydrodynamic zoom simulations are producing large quantities of classical-mass dwarf satellites. The state-of-the-art is to compare their properties (abundance, size, and mass) to the Local Group satellites. However, the significantly larger sample size explored here allows for much more in-depth comparison between observations and simulations. 

Second, we found that the intrinsic shape results were difficult to interpret in the context of the similarity in sizes between late-type and early-type satellites. On the one hand, the similarity of sizes and S\'{e}rsic indices indicated that only mild (if any) structural change occurs when a late-type dwarf becomes an early-type; on the other hand, the thicker intrinsic shapes of early-type dwarfs indicates that some evolution must happen. Simulations that track the evolution in both the size and intrinsic flattening of dwarfs as they are tidally stripped across a range in dwarf mass would help clarify this issue. Alternatively, the rounder shapes could be an effect of the passive fading of a disky young stellar component in a recently-quenched dwarf, bringing out an older, rounder stellar halo component \citep[e.g.][]{kadofong2020}.

Finally, these observations call for more detailed simulations of the transformation of late-type dIrrs to early-type dSphs as they fall into host halos. There is significant existing literature on this transformation in the context of the `tidal stirring' model \citep[e.g.][]{mayer2001a,kazantzidis2011,kazantzidis2017} showing a drastic morphological change between a rotation-supported, disky\footnote{These simulations generally start with a $C/A\sim0.2-0.4$ initial flattening.} dIrr into a dispersion-supported, round dSph is possible. However, this work and others have shown that such a drastic change is generally \textit{not needed}. We have shown here that significant tidal processing simply cannot be part of the transformation since the late-type sizes are so similar to that of the early-types. In essence, in this work we have quantified the starting and finishing states (in terms of size, surface brightness, and intrinsic shape) of the dSph transformation process, and it will be interesting to see if a simulation that incorporates the physics of tidal evolution and quenching can connect the two states.

The observational pathway forward is clearly to establish a robust field sample of isolated dwarfs with well-quantified completeness. We have operated largely under the assumption that the late-type satellite sample largely represents field dwarfs that have not been environmentally processed by a host.  Fortunately, the outlook to procure such a sample is good. The wide-field surveys of both the Vera C. Rubin Observatory and the Roman Space Telescope will allow for the detection \textit{and distance determination} of many field dwarf galaxies. The detection and SBF techniques we have developed in the the ELVES Survey will have natural application to these upcoming surveys.

\section*{Acknowledgements}

Support for this work was provided by NASA through Hubble Fellowship grant \#51386.01 awarded to R.L.B. by the Space Telescope Science Institute, which is operated by the Association of  Universities for Research in Astronomy, Inc., for NASA, under contract NAS 5-26555. J.P.G. is supported by an NSF Astronomy and Astrophysics Postdoctoral Fellowship under award AST-1801921. J.E.G. is partially supported by the National Science Foundation grant AST-1713828. S.G.C acknowledges support by the National Science Foundation Graduate Research Fellowship Program under Grant No. \#DGE-1656466.

Based on observations obtained with MegaPrime/MegaCam, a joint project of CFHT and CEA/IRFU, at the Canada-France-Hawaii Telescope (CFHT) which is operated by the National Research Council (NRC) of Canada, the Institut National des Science de l'Univers of the Centre National de la Recherche Scientifique (CNRS) of France, and the University of Hawaii. This research was made possible through the use of the AAVSO Photometric All-Sky Survey (APASS) \citep{apass}, funded by the Robert Martin Ayers Sciences Fund and NSF AST-1412587

\software{ \texttt{astropy} \citep{astropy} \texttt{sep} \citep{sep} \texttt{imfit} \citep{imfit} \texttt{emcee} \citep{emcee}}

\bibliographystyle{aasjournal}
\bibliography{calib}

\appendix

\section{Photometry of Isolated Field Dwarfs}
\label{app:field}
Here we present our photometry of isolated field dwarfs from the Nearby Galaxy Catalog of \citet{karachentsev2013}. Table \ref{tab:photometry_field} lists the photometry.

\clearpage
\startlongtable
\begin{longrotatetable}
\movetabledown=10mm
\begin{deluxetable}{ccccccccccccc}
\tablecaption{Dwarf Photometry\label{tab:photometry_field}}
\tablehead{
\colhead{Name} & \colhead{Host} & \colhead{RA}  & \colhead{DEC} & \colhead{$M_g$} & \colhead{$M_V$} & \colhead{$\log(M_\star$)}  & \colhead{$\mu_{0,V}$} & \colhead{$r_e$}  & \colhead{$\epsilon$}  & \colhead{ETG?}  & \colhead{filters}  & \colhead{Instrument}  \\ 
\colhead{} & \colhead{} & \colhead{(deg)}  & \colhead{(deg)} & \colhead{(mag)} & \colhead{(mag)} & \colhead{}  & \colhead{(mag/arcsec$^2$)} & \colhead{(pc)}  & \colhead{}  & \colhead{}  & \colhead{}  & \colhead{}  }
\startdata
IC1959  &  FIELD  &  53.302  &  -50.413  &  $-16.05$  &  $-16.26\pm 0.09$  &  $8.27\pm 0.08$  &  $20.63\pm 0.11$  &  $1177.2\pm 56.2$  &  $0.75\pm 0.02$  &  N  &  gr  &  DECALS \\
PGC013294  &  FIELD  &  53.987  &  -45.192  &  $-13.12$  &  $-13.28\pm 0.09$  &  $6.96\pm 0.08$  &  $21.51\pm 0.11$  &  $245.7\pm 13.5$  &  $0.12\pm 0.02$  &  N  &  gr  &  DECALS \\
NGC1705  &  FIELD  &  73.558  &  -53.361  &  $-15.64$  &  $-15.79\pm 0.09$  &  $7.93\pm 0.08$  &  $19.1\pm 0.11$  &  $301.6\pm 14.5$  &  $0.17\pm 0.02$  &  N  &  gr  &  DECALS \\
HIPASSJ0457-42  &  FIELD  &  74.244  &  -42.801  &  $-14.23$  &  $-14.3\pm 0.09$  &  $7.1\pm 0.08$  &  $21.75\pm 0.11$  &  $571.5\pm 28.8$  &  $0.5\pm 0.02$  &  N  &  gr  &  DECALS \\
ESO553-046  &  FIELD  &  81.774  &  -20.678  &  $-14.72$  &  $-14.84\pm 0.09$  &  $7.48\pm 0.07$  &  $20.24\pm 0.11$  &  $348.0\pm 17.2$  &  $0.35\pm 0.02$  &  N  &  gr  &  DECALS \\
HIPASSJ0607-34  &  FIELD  &  91.832  &  -34.204  &  $-15.83$  &  $-15.99\pm 0.09$  &  $8.05\pm 0.07$  &  $19.6\pm 0.11$  &  $634.0\pm 31.0$  &  $0.49\pm 0.02$  &  N  &  gr  &  DECALS \\
ESO121-020  &  FIELD  &  93.979  &  -57.726  &  $-13.57$  &  $-13.64\pm 0.09$  &  $6.87\pm 0.08$  &  $22.95\pm 0.11$  &  $604.7\pm 31.1$  &  $0.29\pm 0.02$  &  N  &  gr  &  DECALS \\
KKH37  &  FIELD  &  101.941  &  80.124  &  $-12.15$  &  $-12.4\pm 0.09$  &  $6.86\pm 0.08$  &  $22.34\pm 0.11$  &  $308.2\pm 16.0$  &  $0.33\pm 0.02$  &  N  &  gr  &  DECALS \\
UGC03600  &  FIELD  &  103.916  &  39.095  &  $-14.94$  &  $-15.16\pm 0.09$  &  $7.87\pm 0.08$  &  $21.88\pm 0.11$  &  $1577.8\pm 80.1$  &  $0.74\pm 0.02$  &  N  &  gr  &  DECALS \\
UGC03698  &  FIELD  &  107.328  &  44.381  &  $-15.16$  &  $-15.42\pm 0.09$  &  $8.08\pm 0.08$  &  $22.17\pm 0.11$  &  $936.4\pm 47.5$  &  $0.33\pm 0.02$  &  N  &  gr  &  DECALS \\
DDO043  &  FIELD  &  112.073  &  40.77  &  $-15.41$  &  $-15.59\pm 0.09$  &  $7.94\pm 0.07$  &  $22.43\pm 0.11$  &  $1347.0\pm 67.1$  &  $0.44\pm 0.02$  &  N  &  gr  &  DECALS \\
NGC2366  &  FIELD  &  112.224  &  69.213  &  $-16.21$  &  $-16.36\pm 0.09$  &  $8.16\pm 0.08$  &  $21.78\pm 0.11$  &  $2465.6\pm 116.4$  &  $0.66\pm 0.02$  &  N  &  gr  &  DECALS \\
DDO046  &  FIELD  &  115.358  &  40.111  &  $-15.57$  &  $-15.7\pm 0.09$  &  $7.85\pm 0.08$  &  $23.44\pm 0.11$  &  $1703.0\pm 84.3$  &  $0.11\pm 0.02$  &  N  &  gr  &  DECALS \\
DDO047  &  FIELD  &  115.482  &  16.802  &  $-15.99$  &  $-16.17\pm 0.09$  &  $8.17\pm 0.08$  &  $22.53\pm 0.11$  &  $4298.9\pm 207.4$  &  $0.63\pm 0.02$  &  N  &  gr  &  DECALS \\
KK65  &  FIELD  &  115.633  &  16.561  &  $-14.5$  &  $-14.69\pm 0.09$  &  $7.6\pm 0.08$  &  $21.64\pm 0.11$  &  $856.7\pm 43.3$  &  $0.64\pm 0.02$  &  N  &  gr  &  DECALS \\
KKH40  &  FIELD  &  116.737  &  51.296  &  $-13.95$  &  $-14.13\pm 0.09$  &  $7.36\pm 0.07$  &  $23.02\pm 0.11$  &  $770.8\pm 41.2$  &  $0.33\pm 0.02$  &  N  &  gr  &  DECALS \\
UGC04115  &  FIELD  &  119.259  &  14.39  &  $-15.17$  &  $-15.34\pm 0.09$  &  $7.8\pm 0.08$  &  $22.33\pm 0.11$  &  $1240.4\pm 61.0$  &  $0.55\pm 0.02$  &  N  &  gr  &  DECALS \\
KDG052  &  FIELD  &  125.985  &  71.031  &  $-11.22$  &  $-11.3\pm 0.09$  &  $5.94\pm 0.08$  &  $24.9\pm 0.11$  &  $398.8\pm 22.4$  &  $0.17\pm 0.02$  &  N  &  gr  &  CFHT \\
DDO052  &  FIELD  &  127.119  &  41.856  &  $-15.18$  &  $-15.44\pm 0.09$  &  $8.11\pm 0.08$  &  $23.05\pm 0.11$  &  $2169.3\pm 108.5$  &  $0.41\pm 0.02$  &  N  &  gr  &  DECALS \\
DDO053  &  FIELD  &  128.536  &  66.177  &  $-13.51$  &  $-13.64\pm 0.09$  &  $7.04\pm 0.08$  &  $23.41\pm 0.11$  &  $907.3\pm 44.6$  &  $0.35\pm 0.02$  &  N  &  gr  &  DECALS \\
AGC182595  &  FIELD  &  132.8  &  27.88  &  $-12.88$  &  $-13.06\pm 0.09$  &  $6.94\pm 0.08$  &  $21.99\pm 0.12$  &  $369.3\pm 21.5$  &  $0.29\pm 0.02$  &  N  &  gr  &  DECALS \\
LSBCD564m08  &  FIELD  &  135.724  &  20.075  &  $-13.13$  &  $-13.37\pm 0.09$  &  $7.23\pm 0.08$  &  $23.71\pm 0.11$  &  $713.6\pm 40.9$  &  $0.05\pm 0.02$  &  N  &  gr  &  DECALS \\
\enddata
\tablecomments{The photometry for the LV field dwarf sample considered in this work.  The full version of the table will be published in the online journal or will be provided upon request to the authors.}
\end{deluxetable}\end{longrotatetable}
\clearpage

\section{Comparison of Photometry with NGVS and NGFS Results}
\label{app:ng_compare}

\begin{figure*}
\includegraphics[width=\textwidth]{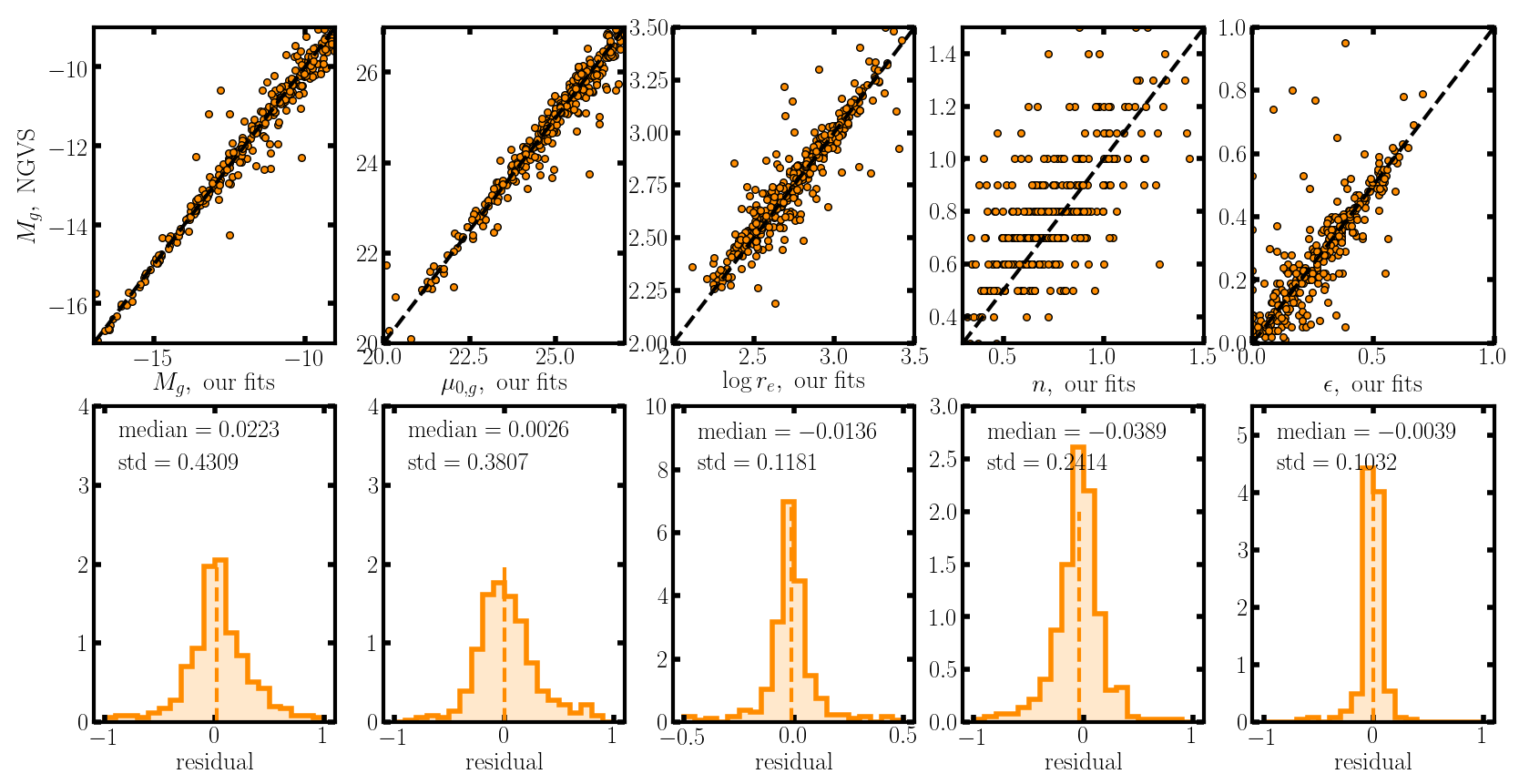}
\caption{Comparison of the NGVS photometry with our photometry of the Virgo sample of dwarfs. Note the NGVS effective radii have been divided by a $\sqrt(1-\epsilon)$ term to account for the fact that the reported effective radii are ``geometric'' radii while the radii reported in this work are all along the major axis. The residual histograms show our photometric measurements subtracted by the NGVS photometry.}
\label{fig:ngvs}
\end{figure*}

\begin{figure*}
\includegraphics[width=\textwidth]{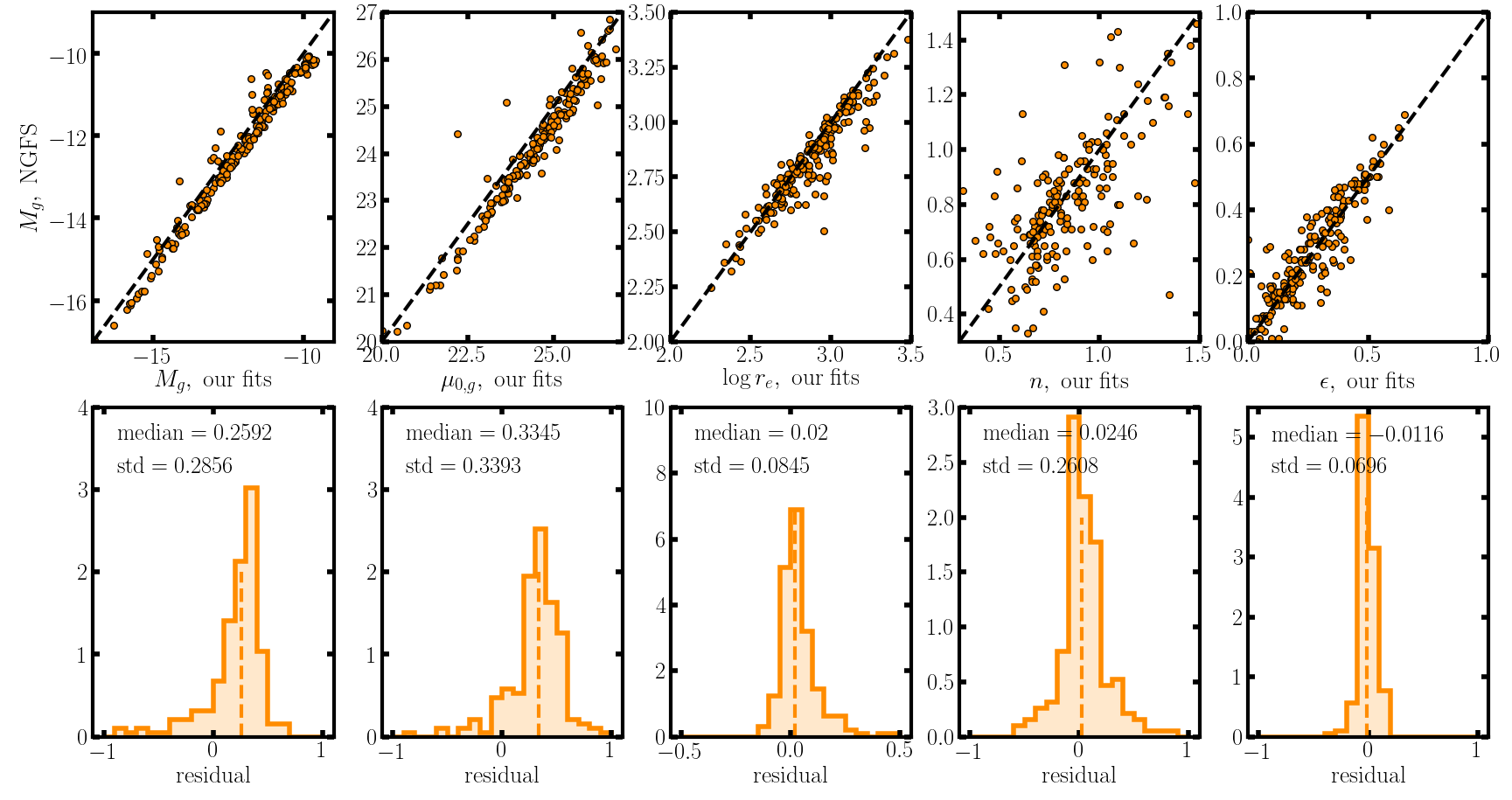}
\caption{Comparison of the NGFS photometry with our photometry of the Fornax sample of dwarfs. Unlike with the NGVS dwarfs, we do not use the same data as the NGFS collaboration but instead use the Dark Energy Survey DECam data. There appears to be a bias in the dwarf magnitudes and surface brightness. }
\label{fig:ngfs}
\end{figure*}

\begin{figure*}
\includegraphics[width=\textwidth]{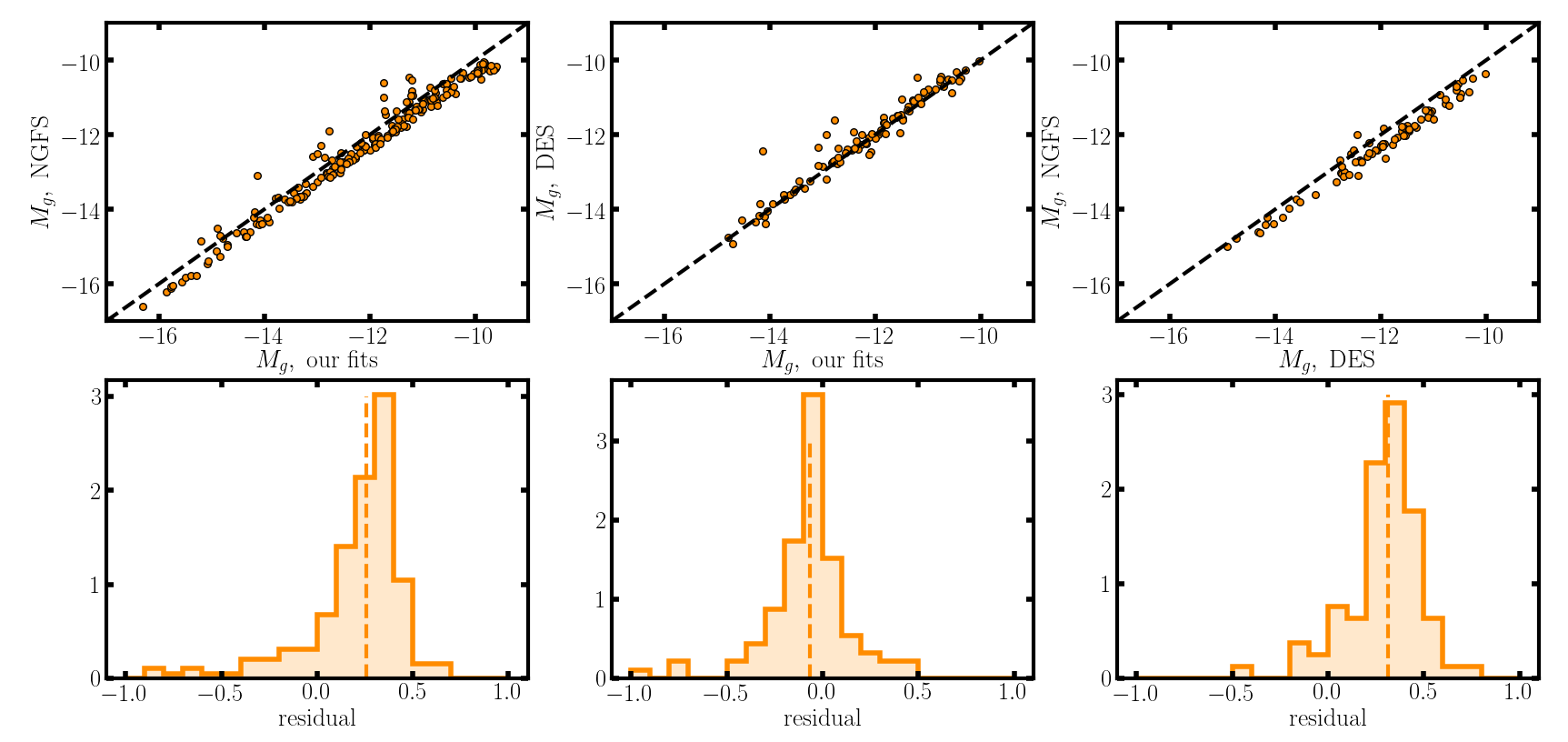}
\caption{Comparison of our photometry of the NGFS dwarfs with both that of the NGFS collaboration and the DES LSBG search of \citet{tanoglidis2020}. }
\label{fig:ngfs2}
\end{figure*}

\begin{figure*}
\includegraphics[width=\textwidth]{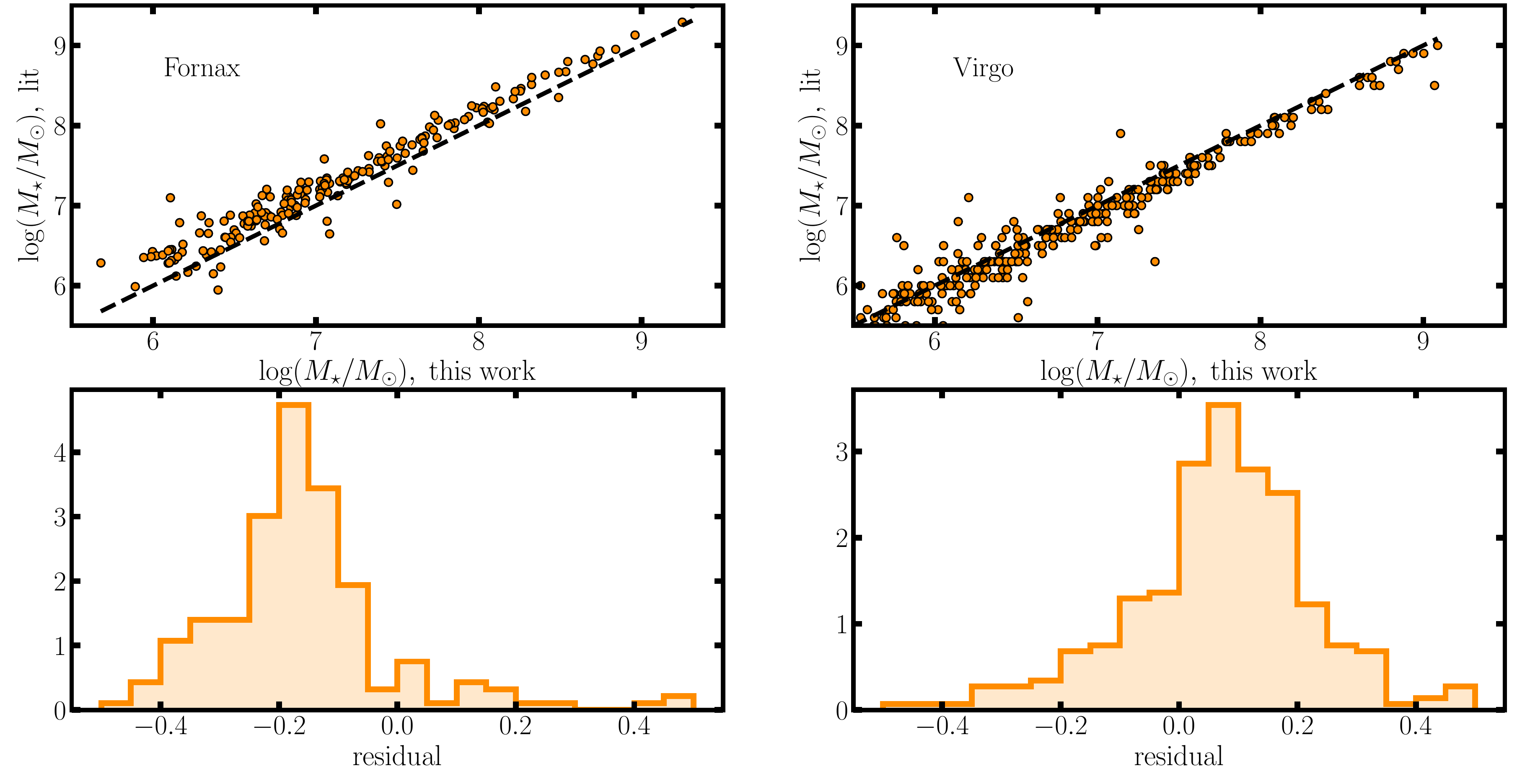}
\caption{Comparison of our stellar mass measurements with those of the NGVS and NGFS groups.}
\label{fig:mstar}
\end{figure*}

In this section, we compare our photometry of the NGVS and NGFS dwarfs with the photometry reported by those collaborations. Figure \ref{fig:ngvs} shows the comparison with the NGVS photometry. The agreement is overall quite remarkable given the different reduction of the data and procedure in fitting S\'{e}rsic profiles. Importantly, the size measurements appear to be unbiased at the $\sim0.01$ dex level, much smaller than the difference we find between average cluster satellite size and LV satellite size.

Figure \ref{fig:ngfs} shows the analogous plot for the Fornax dwarfs. Note that here, unlike with the NGVS dwarfs, we do not use the same data as the NGFS collaboration but instead use the Dark Energy Survey DECam data, as reduced by the DECaLS project. The agreement for size, $n$, and ellipticity is quite good but there is a noticeable bias of $\sim0.3$ mag in the magnitude and surface brightness. Since the size and S\'{e}rsic index are recovered well, the fits themselves are likely robust and it is unlikely to be a difference in sky subtraction. It seems the most likely cause could be a difference in photometric calibration. The image cutouts we use are reduced within the framework of DECaLS which we believe to have a very trustworthy photometric calibration. To explore this some more, in Figure \ref{fig:ngfs2}, we compare our photometry with both that of the NGFS collaboration and that of the low surface brightness galaxy search within DES data by \citet{tanoglidis2020}. The \citet{tanoglidis2020} results agree well with ours and show a similar bias when compared to the NGFS results.

In Figure \ref{fig:mstar}, we compare the stellar masses we measure with those reported by the NGFS and NGVS papers. Stellar masses of the NGVS dwarfs come from \citet{rsj2019} and stellar masses of the NGFS dwarfs come from \citet{eigenthaler2018}. Both comparisons show noticeable biases. It is likely that most, if not all, of the differences can be attributable to different IMF choices in the stellar population modelling. The NGVS results assume a Chabrier IMF \citep{rsj2019}, the NGFS results assume a ``diet'' Salpeter \citep{eigenthaler2018}, while the stellar masses we calculate assume a \citet{kroupa1998} IMF \citep{into2013}. Both the NGFS and NGVS stellar masses are calculated using more than two bands, and thus their estimates are likely more robust. However, ours are calculated in a way that is consistent with the ELVES satellites, which is critical for the comparisons we do in this work. Both comparisons show a bias of about $\sim0.1$ dex which we take as an estimate for the systematic uncertainty stemming from model uncertainties in the color M/L relation, and we include it in all the stellar masses used in the main text. Note, however, that all the comparisons between LV and cluster dwarfs in the main text use stellar masses calculated in the same way: using the color-M/L relations of \citet{into2013} and a \citet{kroupa1998} IMF.

\section{Checks on the Different Filter Systems}
\label{app:filters}
\begin{figure*}
\includegraphics[width=\textwidth]{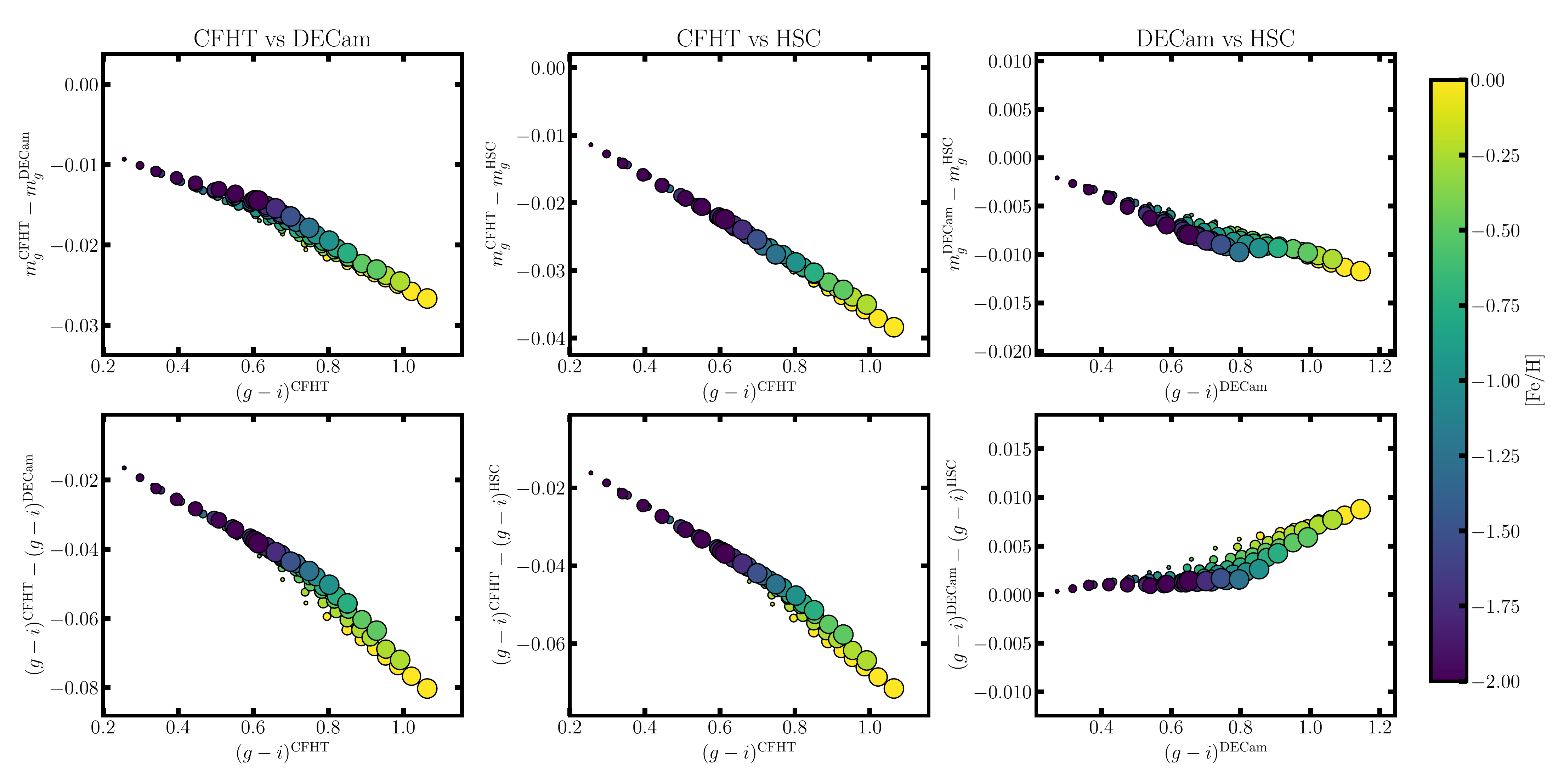}
\caption{A comparison of the different filter systems most used in this work: CFHT/Megacam, Blanco/DECam, and Subaru/HSC. Top row shows the difference in $g$ magnitude for different stellar populations predicted by the MIST isochrones \citep{mist_models} while bottom shows the difference in color. The age of the stellar populations are indicated by point size with the smallest points being 1 Gyr old populations and the largest being 10 Gyr old.}
\label{fig:filters}
\end{figure*}

Here we show that the effect of the myriad of different filter systems considered in this work is generally small compared to the dynamic range in the structural scaling relations. Figure \ref{fig:filters} shows the difference between CFHT/Megacam, Blanco/DECam, and Subaru/HSC filters\footnote{Note that no photometry used in the main text actually comes from HSC observations. HSC observations are used, however, in confirming dwarf satellites via SBF.}. The top row shows the difference in $g$ magnitude while the bottom shows the difference in color. The differences in $g$ magnitude are $\lesssim0.05$ mag which are minuscule compared to the dynamic range of the satellite luminosity considered here ($\sim8$ mag). However, the color differences of $\sim0.05$ mag are significant compared to the color dynamic range ($\sim0.5$ mag). Thus, throughout the paper, we did not heavily analyze the difference colors of the dwarfs. The exception to this is that color is used to determine dwarf stellar masses. However, even a $0.05$ mag systematic bias in the color will change the stellar mass by $\lesssim0.1$ dex which is small compared to both the dynamic range in stellar mass ($\sim3$ dex) and the average random uncertainty in stellar mass ($\sim0.2$ dex).

\section{Dwarf Galaxy Photometry Results}
\label{app:photo_tables}
Here we present the photometry of all the ELVES satellites considered in this work. Table \ref{tab:photometry} shows a sample of the full photometric results. The full table will be published online and will be made available upon request to the authors. Here we list a few details and caveats that should be kept in mind when using this photometry.

We only list photometry for distance confirmed satellites (either with SBF, redshift, or TRGB) and for dwarfs in the mass range $5.5 < \log(M_\star/M_\odot) < 8.5$ with surface brightness $\mu_{0,V} < 26.5$ mag arcsec$^{-2}$. One host (CenA) is missing a few satellites in this mass range for which we could not find appropriate photometry. Thus, this satellite list is not recommended for analyzing the satellite luminosity functions of these groups. Satellite lists more appropriate for that use will be published in future work.

A minority ($<10$\%) of dwarfs were unfortunately located near bright stars or saturation spikes that likely biased the photometry.  These dwarfs are still included in the analysis in this work, but we flag them for the sake of caution in Table \ref{tab:photometry}. Heavily tidally distorted dwarfs are also included in this category.

In analyzing the photometry for satellites of CenA, we found that the measured (extinction corrected) colors of many of the dwarfs were redder ($g-i\sim1$) than expected for low-mass dSphs. It is unclear what is causing this, but since CenA is one of the most extincted hosts in the ELVES sample, we suspect this is some effect of extinction beyond normal reddening (e.g. cirrus). This does not effect the sizes and only minimally affects the integrated luminosities but does affect the stellar mass through the color. Thus, we suspect some of the CenA satellites have somewhat overestimated ($\sim0.3$ dex) stellar masses.

\clearpage
\startlongtable
\begin{longrotatetable}
\movetabledown=10mm
\begin{deluxetable}{ccccccccccccc}
\tablecaption{Dwarf Photometry\label{tab:photometry}}
\tablehead{
\colhead{Name} & \colhead{Host} & \colhead{RA}  & \colhead{DEC} & \colhead{$M_g$} & \colhead{$M_V$} & \colhead{$\log(M_\star$)}  & \colhead{$\mu_{0,V}$} & \colhead{$r_e$}  & \colhead{$\epsilon$}  & \colhead{ETG?}  & \colhead{filters}  & \colhead{Instrument}  \\ 
\colhead{} & \colhead{} & \colhead{(deg)}  & \colhead{(deg)} & \colhead{(mag)} & \colhead{(mag)} & \colhead{}  & \colhead{(mag/arcsec$^2$)} & \colhead{(pc)}  & \colhead{}  & \colhead{}  & \colhead{}  & \colhead{}  }
\startdata
dw0235p3850  &  NGC1023  &  38.976  &  38.836  &  $-13.27$  &  $-13.51\pm 0.15$  &  $7.25\pm 0.12$  &  $21.18\pm 0.1$  &  $532.8\pm 86.4$  &  $0.35\pm 0.08$  &  N$^\dagger$  &  gi  &  CFHT \\
dw0237p3836  &  NGC1023  &  39.414  &  38.6  &  $-11.89$  &  $-12.13\pm 0.17$  &  $6.74\pm 0.12$  &  $23.85\pm 0.13$  &  $547.9\pm 63.0$  &  $0.42\pm 0.15$  &  Y  &  gi  &  CFHT \\
dw0239p3903  &  NGC1023  &  39.843  &  39.055  &  $-9.47$  &  $-9.69\pm 0.49$  &  $5.68\pm 0.15$  &  $25.06\pm 0.28$  &  $335.5\pm 44.0$  &  $0.52\pm 0.06$  &  Y  &  gi  &  CFHT \\
dw0239p3902  &  NGC1023  &  39.946  &  39.047  &  $-9.51$  &  $-9.82\pm 0.09$  &  $6.02\pm 0.15$  &  $24.72\pm 0.07$  &  $284.5\pm 15.8$  &  $0.51\pm 0.03$  &  Y  &  gi  &  CFHT \\
dw0239p3926  &  NGC1023  &  39.832  &  39.434  &  $-11.87$  &  $-12.11\pm 0.14$  &  $6.72\pm 0.12$  &  $26.17\pm 0.1$  &  $1129.4\pm 103.8$  &  $0.25\pm 0.05$  &  Y  &  gi  &  CFHT \\
dw0240p3903$^*$  &  NGC1023  &  40.154  &  39.059  &  $-15.68$  &  $-15.98\pm 0.09$  &  $8.41\pm 0.12$  &  $21.97\pm 0.09$  &  $2047.4\pm 0.0$  &  $0.22\pm 0.02$  &  N  &  gi  &  CFHT \\
dw0240p3922$^*$  &  NGC1023  &  40.165  &  39.379  &  $-13.41$  &  $-13.52\pm 0.07$  &  $6.84\pm 0.11$  &  $22.45\pm 0.05$  &  $642.2\pm 53.4$  &  $0.53\pm 0.07$  &  N  &  gi  &  CFHT \\
dw0932p2127  &  NGC2903  &  143.203  &  21.466  &  $-14.03$  &  $-14.37\pm 0.08$  &  $7.87\pm 0.13$  &  $22.38\pm 0.13$  &  $728.9\pm 30.3$  &  $0.14\pm 0.02$  &  Y  &  gr  &  CFHT \\
dw0930p1959  &  NGC2903  &  142.554  &  19.991  &  $-13.12$  &  $-13.34\pm 0.09$  &  $7.13\pm 0.13$  &  $23.61\pm 0.11$  &  $556.5\pm 32.0$  &  $0.07\pm 0.02$  &  N  &  gr  &  DECALS \\
dw0933p2030  &  NGC2903  &  143.307  &  20.515  &  $-12.39$  &  $-12.65\pm 0.1$  &  $6.99\pm 0.13$  &  $22.53\pm 0.12$  &  $443.9\pm 28.3$  &  $0.3\pm 0.03$  &  Y  &  gr  &  DECALS \\
dw0936p2135  &  NGC2903  &  144.089  &  21.599  &  $-11.93$  &  $-12.19\pm 0.1$  &  $6.78\pm 0.13$  &  $22.74\pm 0.13$  &  $341.4\pm 23.6$  &  $0.2\pm 0.03$  &  N  &  gr  &  DECALS \\
dw0930p2143  &  NGC2903  &  142.667  &  21.724  &  $-10.81$  &  $-10.95\pm 0.09$  &  $5.98\pm 0.12$  &  $23.3\pm 0.06$  &  $270.5\pm 21.5$  &  $0.54\pm 0.02$  &  N  &  gr  &  CFHT \\
dw0932p1952  &  NGC2903  &  143.099  &  19.875  &  $-10.44$  &  $-10.78\pm 0.12$  &  $6.43\pm 0.15$  &  $24.17\pm 0.17$  &  $309.0\pm 30.6$  &  $0.38\pm 0.05$  &  Y  &  gr  &  DECALS \\
LVJ1218+4655  &  NGC4258  &  184.547  &  46.917  &  $-12.8$  &  $-12.92\pm 0.02$  &  $6.71\pm 0.1$  &  $21.85\pm 0.05$  &  $562.2\pm 20.4$  &  $0.75\pm 0.04$  &  N  &  gr  &  CFHT \\
ngc4258-df6  &  NGC4258  &  184.776  &  47.73  &  $-11.13$  &  $-11.35\pm 0.16$  &  $6.36\pm 0.14$  &  $24.43\pm 0.1$  &  $484.4\pm 44.2$  &  $0.12\pm 0.03$  &  Y  &  gr  &  CFHT \\
kdg101  &  NGC4258  &  184.787  &  47.09  &  $-14.01$  &  $-14.31\pm 0.1$  &  $7.73\pm 0.13$  &  $23.08\pm 0.16$  &  $887.9\pm 60.2$  &  $0.32\pm 0.03$  &  Y  &  gr  &  CFHT \\
dw1220p4649  &  NGC4258  &  185.229  &  46.83  &  $-10.39$  &  $-10.66\pm 0.14$  &  $6.23\pm 0.15$  &  $25.21\pm 0.08$  &  $422.2\pm 23.6$  &  $0.23\pm 0.03$  &  Y  &  gr  &  CFHT \\
dw1217p4724  &  NGC4258  &  184.459  &  47.409  &  $-16.6$  &  $-16.89\pm 0.02$  &  $8.73\pm 0.1$  &  $20.22\pm 0.04$  &  $1877.2\pm 17.1$  &  $0.71\pm 0.0$  &  N  &  gr  &  CFHT \\
dw1223p4739  &  NGC4258  &  185.942  &  47.659  &  $-11.16$  &  $-11.42\pm 0.09$  &  $6.46\pm 0.14$  &  $24.88\pm 0.07$  &  $555.8\pm 80.9$  &  $0.38\pm 0.14$  &  Y  &  gr  &  CFHT \\
dw1234p2531  &  NGC4565  &  188.601  &  25.522  &  $-13.66$  &  $-13.97\pm 0.03$  &  $7.65\pm 0.13$  &  $23.33\pm 0.03$  &  $1074.2\pm 28.2$  &  $0.53\pm 0.01$  &  Y  &  gr  &  CFHT \\
dw1235p2551  &  NGC4565  &  188.895  &  25.85  &  $-16.86$  &  $-17.13\pm 0.01$  &  $8.79\pm 0.1$  &  $21.05\pm 0.0$  &  $2004.3\pm 8.0$  &  $0.66\pm 0.0$  &  N  &  gr  &  CFHT \\
dw1237p2602  &  NGC4565  &  189.255  &  26.036  &  $-12.33$  &  $-12.56\pm 0.06$  &  $6.85\pm 0.1$  &  $22.22\pm 0.31$  &  $457.8\pm 33.6$  &  $0.49\pm 0.01$  &  N  &  gr  &  CFHT \\
\enddata
\tablecomments{The photometry for the main LV satellite dwarf sample considered in this work. Dwarfs marked with a $^*$ had significant contamination from bright nearby stars or other galaxies, and the photometry might be biased. Dwarf marked with $^\dagger$ had ambiguous ETG/LTG visual classification and were classified based on the dwarf color and magnitude. The full version of the table will be published in the online journal or will be provided upon request to the authors. Sources are those as listed in Table 
ef{tab:hosts}}
\end{deluxetable}\end{longrotatetable}
\clearpage

\section{Estimating the Uncertainty in the Dwarf Photometry}
\label{app:uncertainty}

\begin{figure*}
\includegraphics[width=\textwidth]{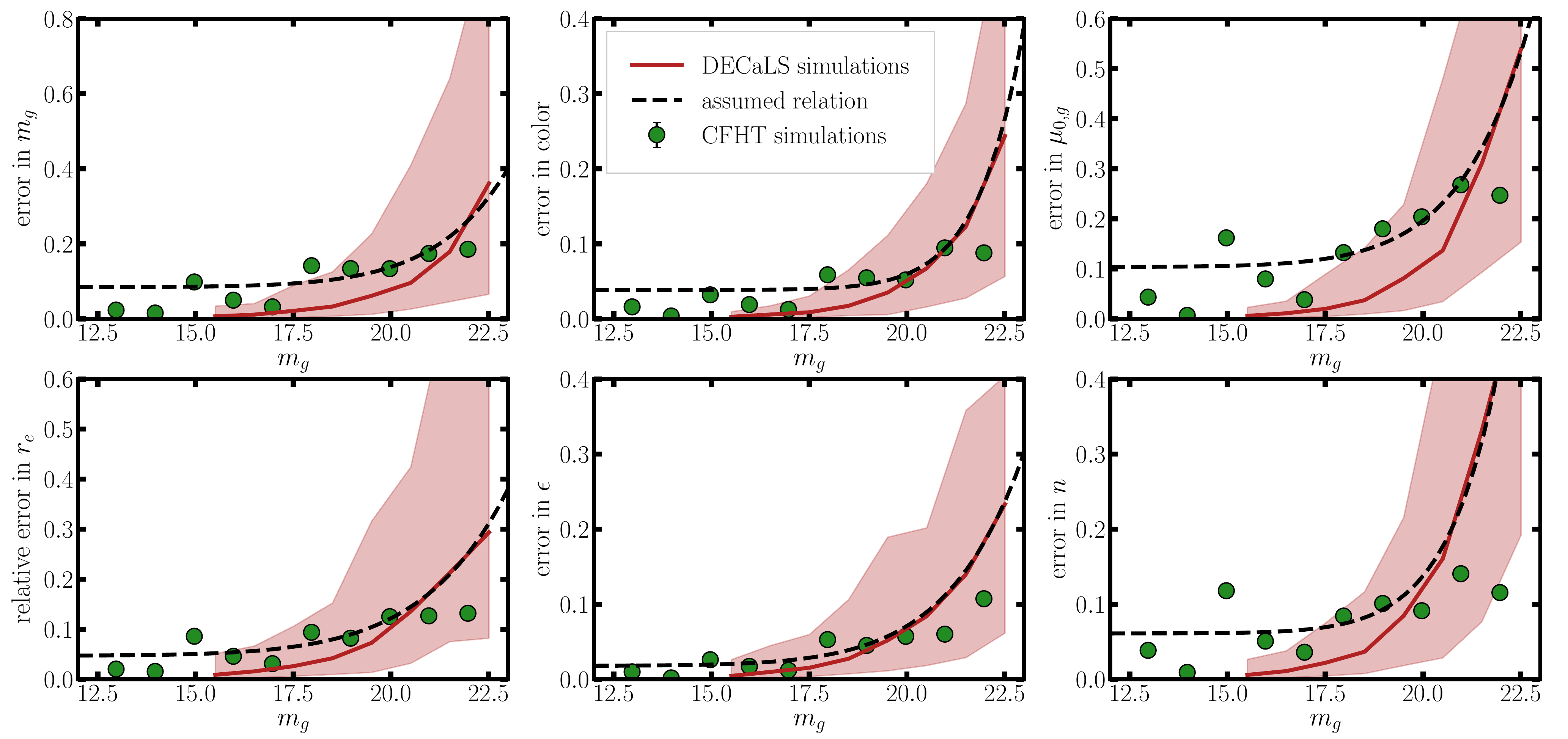}
\caption{The functions that we use in this paper that give estimated error for the different structural quantities as a function of apparent $g$ magnitude. We show both the average error in the recovery of the artificial galaxy simulations of \citet{carlsten2020a} and also the new image simulations that we do for this work in the DECaLS data. Note that the CFHT image simulations account for the uncertainty due to sky subtraction while the DECaLS simulations do not. Due to this, the estimated photometric errors are greater for the CFHT simulations at bright magnitudes even though the DECaLS data are much shallower. The dashed black curve shows the relation we use to estimate the photometric uncertainties for the DECaLS dwarfs in this paper which essentially fits the DECaLS simulations at faint magnitudes but uses the error floor of the CFHT simulations at brighter magnitudes.}
\label{fig:photo-errors}
\end{figure*}

Here we give more details on how we estimate the uncertainty on the photometric measurements of the dwarf satellites. In \citet{carlsten2020a}, we use image simulations of artificial galaxies to estimate the uncertainty in the derived photometry from dwarfs, on a dwarf-by-dwarf basis, in the CFHT datasets. We use those errors here when available. For the rest of the dwarfs considered here (which almost entirely were detected in DECaLS), we perform new image simulations to estimate the uncertainties in the derived structural quantities. We take a random DECaLS tract ($3600\times3600$ pixel region) from each of the DECaLS hosts and use them as backgrounds for the image simulations. We inject dwarfs using S\'{e}rsic profiles with a range in apparent magnitude into the data and quantify how well we are able to recover the input structural quantities. The dwarfs are given sizes based on a luminosity-size relation roughly what we find in this work (we take the distance to each dwarf as a random value between 5 and 10 Mpc). Dwarf color is uniformly drawn from the range [0.2,0.7], ellipticity is drawn uniformly from the range [0,0.4], and S\'{e}rsic index is drawn from Gaussian with $(\mu,\sigma)=(0.75,0.2)$. We then fit S\'{e}rsic profiles to the artificial galaxies and compare the recovered parameters with the input. 

Figure \ref{fig:photo-errors} show the average recovery error in various parameters as a function of apparent magnitude for both the CFHT simulations of \citet{carlsten2020a} and the DECaLS simulations we do here. The error in recovery increases for fainter input galaxies, as expected. The CFHT simulations included the effect of sky subtraction errors as the artificial galaxies were injected prior to sky subtraction and coaddition. This is why the recovery errors are greater for the CFHT simulations at bright magnitudes even though the DECaLS data is much shallower. We use these errors in the recovery as estimates of the uncertainty in the photometry of the real DECaLS dwarfs. The function we use to estimate the uncertainty as a function of the apparent $g$ magnitude is shown as the dashed black line which fits the DECaLS simulations at faint magnitudes but uses the error floor of the CFHT simulations at brighter magnitudes.

\section{Color-Magnitude Relation of LV Satellites Dwarfs}
\label{app:morph}

\begin{figure}
\includegraphics[width=0.5\textwidth]{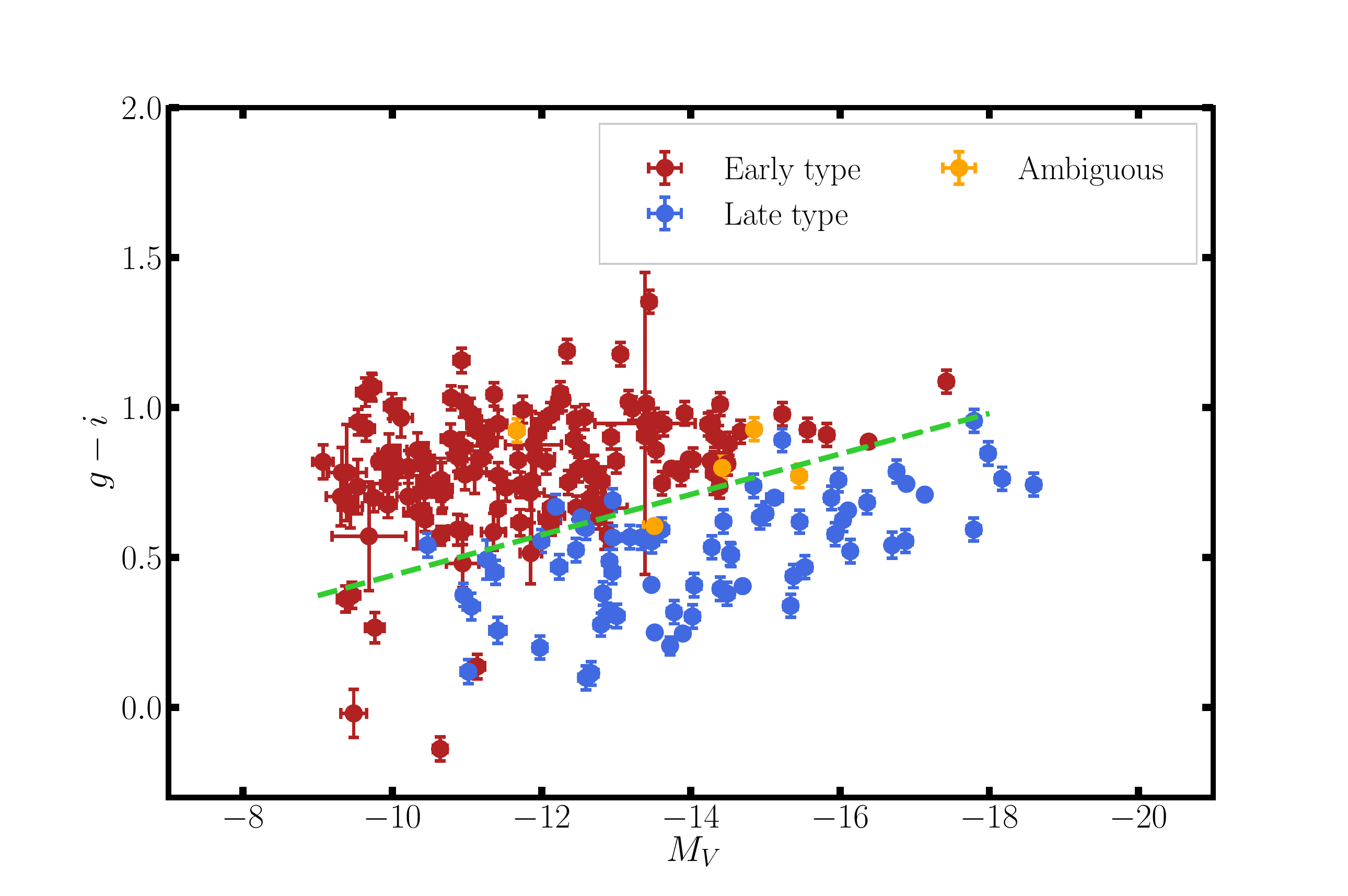}
\caption{Color-magnitude relation of ELVES dwarfs split by the visual morphology classification. There is a clear separation in this space with late-type dwarfs being significantly bluer. The dwarfs that were ambiguous in their visual classification are indicated. The line is the curve given in Section \ref{sec:morph} that was chosen by eye to split the two populations. The two ETGs with very blue ($g-i\lesssim0$) colors are dw0242p3838 and KKs57. dw0242p3838's photometry was heavily corrupted by a bright foreground star while we suspect KKs57's blue color is likely due to photometric calibration problems.}
\label{fig:cmr}
\end{figure}

In this section, we demonstrate that the visual classification of dwarfs into early- and late-type gives essentially the same results as a cut in color-magnitude space. This is shown in Figure \ref{fig:cmr}. The late-type and early-type dwarfs clearly separate in this space. The dividing line between the two populations is given in Sect \ref{sec:morph}. The visually ambiguous dwarfs are classified simply by whether they are above or below this line.

\end{document}